\documentclass[english,preprint,aps,prd,showpacs,superscriptaddress,nofootinbib,tightenlines]{revtex4}
\usepackage[T1]{fontenc}
\usepackage[latin9]{inputenc}
\usepackage{float}
\usepackage{amsmath}
\usepackage{esint}
\usepackage{ulem}
\usepackage{amsfonts}
\usepackage{multirow}
\usepackage{mathrsfs}
\usepackage{graphicx}
\usepackage{amsmath}
\usepackage{amssymb}
\usepackage{bm}
\usepackage{bbm}
\usepackage{xcolor}
\usepackage{babel}
\usepackage{slashed}
\newcommand {\bseq}{\begin{subequations}}
\newcommand {\eseq}{\end{subequations}}
\newcommand {\pvint}[2]{{\int\!\!\!\!\!\!-}_{\!\!\!\!#1}^{#2}}
\newcommand*{\rom}[1]{\uppercase\expandafter{\romannumeral #1\relax}}

\def\OMIT#1{}

\newcommand{\nn}{\nonumber}

\newcommand{\beq}{\begin{equation}}
\newcommand{\eeq}{\end{equation}}
\newcommand{\bqa}{\begin{eqnarray}}
\newcommand{\eqa}{\end{eqnarray}}

\graphicspath{{./figs/}}

\begin{document}

\title{Solving the Bars-Green equation for moving mesons in two-dimensional QCD}

\author{Yu Jia\footnote{jiay@ihep.ac.cn}}
\affiliation{Institute of High Energy Physics and Theoretical Physics Center for Science Facilities, Chinese Academy of Sciences, Beijing 100049, China\vspace{0.2cm}}
\affiliation{School of Physics, University of Chinese Academy of Sciences,
Beijing 100049, China\vspace{0.2cm}}
\affiliation{Center
for High Energy Physics, Peking University, Beijing 100871,
China\vspace{0.2cm}}

\author{Shuangran Liang\footnote{liangsr@ihep.ac.cn}}
\affiliation{Institute of High Energy Physics and Theoretical Physics Center for Science Facilities, Chinese Academy of Sciences, Beijing 100049, China\vspace{0.2cm}}
\affiliation{School of Physics, University of Chinese Academy of Sciences,
Beijing 100049, China\vspace{0.2cm}}

\author{LiuJi Li\footnote{liuji.li@icloud.com}}
\affiliation{Institute of High Energy Physics and Theoretical Physics Center for Science Facilities, Chinese Academy of Sciences, Beijing 100049, China\vspace{0.2cm}}

\author{Xiaonu Xiong\footnote{x.xiong@fz-juelich.de}}
\affiliation{Institute for Advanced Simulation,
Institut f\"ur Kernphysik and J\"ulich Center for Hadron Physics,
Forschungszentrum J\"ulich, D-52425 J\"ulich, Germany}
\affiliation{Istituto Nazionale di Fisica Nucleare, Sezione di Pavia, Pavia, 27100, Italy}

\date{\today}

\begin{abstract}
The two-dimensional QCD in the large $N$ limit, generally referred to as the 't Hooft model,
is numerically investigated in the axial gauge in a comprehensive manner.
The corresponding Bethe-Salpeter equation for a bound $q\bar{q}$ pair, originally derived
by Bars and Green in 1978, was first numerically tackled by Li and collaborators in late 1980s,
yet only for the {\it stationary} mesons.
In this paper, we make further progress by numerically
solving the Bars-Green equation for {\it moving} mesons, ranging from
the chiral pion to charmonium.
By choosing several different quark masses, we computed the corresponding
quark condensates, meson spectra and their decay constants for a variety of meson momenta,
and found satisfactory agreement with their counterparts obtained using light-cone gauge,
thus numerically verified the gauge and Poincar\'{e} invariance of the 't Hooft model.
Moreover, we have explicitly confirmed that, as the meson gets more and more boosted,
the large component of the Bars-Green wave function indeed approaches the corresponding 't Hooft
light-cone wave function, while the small component of the wave function
rapidly fades away.
\end{abstract}


%%%%%%%%%%%%%%%%%%%%%%%%%%%%%%%%%%%%%%%%%%%%%%%%%%%%%%%%%%%%
%%%%%%%%%%%%%%%%%%%%%%%%%%%%%%%%%%%%%%%%%%%%%%%%%%%%%%%%%%%%

%11.10.Kk Field theories in dimensions other than four (see also 04.50.-h Higher-dimensional gravity and other theories of gravity; %04.60.Kz Lower dimensional models; minisuperspace models in general relativity and gravitation)

%11.10.St Bound and unstable states; Bethe-Salpeter equations

%11.15.Pg Expansions for large numbers of components (e.g., 1/Nc expansions)

%11.30.Cp Lorentz and Poincare invariance

%12.38.-t Quantum chromodynamics (for quarks, gluons, and QCD in nuclear reactions, see 24.85.+p)

%12.38.Lg Other nonperturbative calculations

%%%%%%%%%%%%%%%%%%%%%%%%%%%%%%%%%%%%%%%%%%%%%%%%%%%%%%%%%%%%
%%%%%%%%%%%%%%%%%%%%%%%%%%%%%%%%%%%%%%%%%%%%%%%%%%%%%%%%%%%%

\maketitle

\section{Introduction}

Two-dimensional ${\rm QCD}$ (hereafter ${\rm QCD}_2$) has long served as a valuable toy model to
mimic some essential dynamical features of strong interaction in the real world.
A gratifying feature of this theory is that, due to the absence of transverse degree of freedom, the gluon is no longer a dynamical degree of freedom
(at least in non-covariant gauges), but merely provides a linear color Coulomb potential, with the quark confinement as an almost trivial outcome.
Despite this great simplification, ${\rm QCD}_2$ still constitutes a rather nontrivial quantum field theory,
which contains rich hadron phenomenology for mesons and baryons.
There have been some numerical explorations of the ${\rm QCD}_2$ with finite $N$
based on the first-principle approaches,
{\rm e.g.}, from the lattice Monte Carlo simulations~\cite{Hamer:1981yq,Berruto:2002gn}
and from discretized light-cone quantization~\cite{Hornbostel:1988fb}.

The $1/N$ expansion is a powerful and indispensable
arsenal to tackle the nonperturbative dynamics of QCD~\cite{tHooft:1973alw,Witten:1979kh,Coleman:1985}.
As first exemplified in a 1974 seminal paper by 't Hooft~\cite{'tHooft:1974hx},
thanks to the dominance of the planar diagrams, ${\rm QCD}_2$ in the large $N$ limit
(hereafter abbreviated as the 't Hooft model) indeed becomes much more tractable.
The limit of infinite number of colors is in the following sense:
%---------------------------
\begin{equation}
%---------------------------
N\to\infty,\qquad \lambda\equiv {g^2 N\over 4\pi} \; {\rm fixed},\qquad m_f\gg g\sim {1\over \sqrt{N}},
%---------------------------
\end{equation}
%---------------------------
where $g$ is the strong coupling constant in ${\rm QCD}_2$, which carries the mass dimension one,
and $\lambda$ is often referred to as 't Hooft coupling constant.
The first two conditions are standard large $N$ assumptions.
The last requirement, that the quark masses, $m_f$, are much greater than the gauge coupling,
is usually referred to as the weak coupling regime of the ${\rm QCD}_2$.
The bulk of investigation on 't Hooft model has been mainly concentrating on
this particular regime.

Employing the light-cone gauge $A^{a}_+(x)=0$ and invoking the large-$N$ limit,
't Hooft showed that mass spectra of an infinite tower of mesonic states
can be inferred from the following integral equation:
%-------------------
\beq
%-------------------
\left(M_n^2-\frac{m^2_1 - 2\lambda}{x} -\frac{m^2_2-2\lambda}{1-x} \right)\phi^{(n)} (x) = -
2\lambda \pvint{0}{1}\frac{dy}{(x-y)^{2}}\phi^{(n)}(y),
%-------------------
\label{tHooft:equation}
\eeq
%-------------------
which is nothing but the light-cone Bethe-Salpeter equation for a relativistic $q\bar{q}$ bound state.
$m_{1,2}$ are quark (antiquark) bare masses, $M_n$ denotes the mass of the $n$-th excited mesonic
state ($n$ is the principal quantum number to characterize a meson living on a string),
and $\phi^{(n)}(x)$ signifies the corresponding light-cone wave function,
with $x\in [0,1]$ representing the fraction of the light-cone momentum carried
by the quark with respect to that by the meson.
The symbol $\pvint{}{}$  implies that a principle-value prescription
is exerted to eliminate the infrared divergence that occurs at $y\to x$.
In general, Eq.~(\ref{tHooft:equation}) is not admissible to an analytic solution,
yet can only be solved using numerical recipes.

Intriguingly, the discrete meson spectrum determined from (\ref{tHooft:equation})
exhibits the Regge trajectory.
Specifically speaking, for highly excited states ($n\gg 1$),
the squared meson masses are well described by
%-------------------
\beq
%-------------------
M_n^2 = 2\pi^2 \lambda n + \left(m_1^2+m_2^2-4\lambda\right) \ln n+ {\cal O}(n^0).
%-------------------
\label{Regge:trajectory}
\eeq
%-------------------

In 't Hooft's pioneering work, some essential properties that closely resemble the ordinary QCD,
such as color confinement and Regge trajectory, have already been revealed.
Afterwards there have been extensive investigations on various ``phenomenological''
aspects of the large-$N$ ${\rm QCD}_2$, {\it e.g.},
hadron decay/scattering amplitudes, current correlators, form factors, (naive) asymptotic freedom
and parton model, fragmentation functions, Pomeron, (generalized) parton distribution functions,
quark-hadron duality, and many more else~\cite{Callan:1975ps,Einhorn:1976uz,Brower:1978wm,Lenz:1991sa,Burkardt:2000uu,Glozman:2012ev}.
Apart from these work, a very remarkable feature of this model has also been uncovered in the mid 1980s:
the non-vanishing quark condensate, the spontaneous breaking of chiral symmetry, and
the clarification of (quasi-)Goldstone mode~\cite{Witten:1978qu,Zhitnitsky:1985um,Li:1986gf}.

Historically, most aforementioned features of the 't Hooft model have been deduced
by utilizing the light-cone gauge (often peppered with the light front (LF) quantization).
The greatest advantage of this procedure is that, it allows to yield the manifest boost-invariant
bound-state equation, (\ref{tHooft:equation}), and generates compact expressions for various physical quantities.
Nevertheless, there also exist some disadvantages inherent to this approach, {\it e.g.},
the mechanism initiating chiral symmetry breaking becomes obscure, due to the perturbative nature
of the vacuum in LF quantization.

Interestingly, there also exists an alternative perspective to tackle ${\rm QCD}_2$,
that is, by imposing the axial gauge ($A^a_1(x)=0$) condition in the ordinary equal-time quantization.
Despite its technical complication, this approach does possess some notable virtues. For example,
unlike the light front quantization, the equal-time quantization
accommodates a nontrivial vacuum state, which makes the study of the
spontaneous chiral symmetry breaking much more transparent.
Moreover, this approach has a natural connection to the familiar
constituent quark model,
in analogy with the intimate connection between the light-cone gauge
and the parton model.

In 1978, by quantizing 't Hooft model in the axial gauge, Bars and Green presented a formal proof that
Poincar\'{e} algebra does close in the color-singlet channel~\cite{Bars:1977ud}.
They further derived the axial gauge Bethe-Salpeter equations for mesons in an arbitrary frame.
The resulting relativistic bound-state equation (hereafter dubbed Bars-Green equation) does look
much more sophisticated than its light-cone counterpart, Eq.~(\ref{tHooft:equation}).
As is widely known, it is highly nontrivial to conduct the Lorentz boost for a
bound-state wave function constructed in the equal-time quantization procedure~\cite{Brodsky:1997de}.
It was anticipated that Bars-Green equation must preserve Poincar\'{e}  invariance, {\it i.e.},
the meson mass should not rely at all on which Lorentz frame one is carrying out the measurement,
which is clearly a rudimentary requirement for any sensible theory.
A special and gratifying situation is when a meson is viewed in the infinite momentum frame (IMF),
the Bars-Green equation can be proven to exactly reduce to the 't Hooft equation~\cite{Bars:1977ud}.
Nevertheless, it remains an analytic challenge to prove that Bars-Green equation does
preserve Poincar\'{e}  invariance in any finite momentum reference frame.

In general, it is impossible to solve the Bars-Green equations in an analytic fashion.
The numerical investigation of these equations were pioneered by Li and collaborators
in late 1980s~\cite{Li:1987hx}, but only for the mesons in the zero-momentum frame.
They indeed confirmed that the calculated meson spectra using axial gauge
agreed with what were found by solving the 't Hooft equation.

The aim of this work is to extend the earlier investigation in \cite{Li:1987hx},
by numerically solving the Bars-Green equations for a generic moving meson, with hadron species ranging from
the chiral pion to heavy quarkonium.
Our primary goal is to {\it numerically} validate the Poincar\'{e} invariance of the Bars-Green equations.
Moreover, we wish to quantitatively assert that, to which extent when a meson gets boosted,
the Bars-Green wave function would resemble the corresponding 't Hooft wave function to a decent degree.

It is worth mentioning that, the Bars-Green wave function is intimately related
to the so-called quasi-distributions in ${\rm QCD}_4$, which have received lots of attention in past few years.
The quasi-distributions, a set of instantaneous yet spatially non-local correlators,
was recently introduced by Ji as a proxy to help extract the light-cone distributions,
which can be directly computed by lattice simulation~\cite{Ji:2013dva}.
One of the key properties of the quasi-distributions is that, when boosted to the IMF, they will
reduce to their light-cone cousins, {\it e.g.}, parton distribution functions and light-cone distribution amplitudes.
We wish that our comprehensive numerical study of Bars-Green wave functions
will lend some guidance on quantitatively understanding of the properties of
quasi-distributions in realistic QCD.

The paper is structured as the following.
%----------------------------
In Section~\ref{Derivation:BG:equation}, we present a relatively succinct, yet self-contained
review on the course of arriving at the Bars-Green equation in the 't Hooft model.
We also add more details in illustrating how does the Bars-Green equation reduce to 't Hooft equation in the IMF.
%----------------------------
In Section~\ref{Chiral:condensate:decay:constants}, we investigate the renormalized chiral condensates with a variety of quark masses
in the axial gauge, and compare with the respective values obtained in the light-cone gauge.
We also present the analytical formulas for the decay constants of the even- and odd-parity mesons.
%----------------------------
In Section~\ref{Numerical:recipes}, we briefly describe our numerical strategies in solving the 't Hooft equation
and Bars-Green equations.
%----------------------------
In Section~\ref{Numerical:Results:Collections}, we then present comprehensive numerical studies of the mass spectra and Bars-Green wave functions for a variety of meson species: chiral $\pi$, physical pion, a fictitious strangeonium, and charmonium,
for each of which several different meson momenta are chosen.
We also examine how fast the Bars-Green wave function for a highly boosted meson converges to the respective 't Hooft wave function.
We conclude this Section by examining the frame-independence of the decay constants.
%----------------------------
Finally we summarize in Section~\ref{Summary}.
%----------------------------

\section{Review of the Bars-Green formalism}
\label{Derivation:BG:equation}

In this section, our main goal is to sketch some key intermediate steps in deriving the Bars-Green equation in axial gauge.
Nothing in this section is really new, and the purpose of including this section is mainly for the sake of completeness.
We will follow the Feynman diagramatic approach to derive the BG equation. It is worth noting that, there also exists an elegant alternative way to derive the same equations from the Hamiltonian approach.

We start from the Lagrangian of the 1$+$1-dimensional QCD with the $SU(N)$ color gauge symmetry:
%-------------------
\begin{equation}
%-------------------
\mathcal{L}_{{\rm QCD}} = -\frac{1}{4} F^a_{\mu\nu} F^{a\mu\nu} + \sum_f \bar{q}_f (i D\!\!\!\!/ - m_f )q_f,
\end{equation}
%-------------------
where the gluon field strength $F^a_{\mu\nu} = \partial_\mu  A^a_{\nu}- \partial_{\nu}A^a_{\mu} + g f^{abc} A^b_{\mu}A^c_{\nu}$,
the gauge covariant derivative $D_\mu  = \partial_\mu + i g A^a_\mu T^a $, with $T^a$ being the color $SU(N)$ generator in fundamental representation and $a$ running from 1 to $N^2-1$.
We adopt the Dirac-Pauli representation for the $\gamma$-matrices:
$\displaystyle \gamma^0 = \sigma^3,
\;\gamma^1 = i\sigma^2,\;\gamma^5 = \gamma^0 \gamma^1 = \sigma^1$.

Throughout this work, we are imposing the axial gauge condition $A^a_1(x)=0$~\footnote{In ${\rm QCD}_2$,
the axial gauge is equivalent to the Coulomb gauge, and these two terms are often used interchangeably.}.
Like in the light-cone gauge,
the nonlinear term in $F_{01}^a$, the major characteristic complication of QCD,
simply drops out in the axial gauge. Moreover, $A^a_0$ is no longer a dynamical variable, instead
can be expressed in term of a quark current through the Euler-Lagrange equation.
Hence, in the canonical Hamiltonian form, the gluon field $A^a_0(x)$
has been completely eliminated, whose effects are fully encoded in the instantaneous, yet
spatially-nonlocal current-current interaction.
As a common practice, the current-current interaction can often be
simulated by a gluon propagator in the axial gauge:
%-------------------
\begin{equation}
%-------------------
D^{ab}_{\alpha\beta}(x^\mu) = -{i\over 2}\delta^{ab} \delta_{\alpha 0}\delta_{\beta 0} |x^1|\delta(x^0),
%-------------------
\end{equation}
%-------------------
with the only survivor from the $00$-component, and $x^\mu=(x^0,x^1)$.
It can be immediately identified with the {\it instantaneous linear}
Coulomb potential. It is instructive to rewrite it as a Fourier integral:
%-------------------
\begin{equation}
%-------------------
D^{ab}_{\alpha\beta}(x^\mu) = \delta^{ab} \delta_{\alpha 0}\delta_{\beta 0}
\int_{-\infty}^{\infty} {d k^0\over 2\pi} e^{-i k^0 x^0}
\pvint{-\infty}{\infty} {d k^1\over 2\pi} e^{i k^1 x^1}{i\over (k^1)^2}.
%-------------------
\end{equation}
%-------------------
The momentum-space gluon propagator only depends upon the spatial component of $k^\mu$,
reflecting that $A^a_0$ in the axial gauge is a non-propagating degree of freedom.
Moreover, due to the singular behavior of the integrand near $k^1\to 0$,
one must introduce a proper prescription to make the above Fourier integral well-defined.
This is the origin of the ubiquitous occurrence of the
principle value prescription in two-dimensional gauge theory.

\subsection{Mass-Gap Equation}

 %-------------------------
 \begin{figure}[htbp]
 	\centering
 	\includegraphics[width=\textwidth]{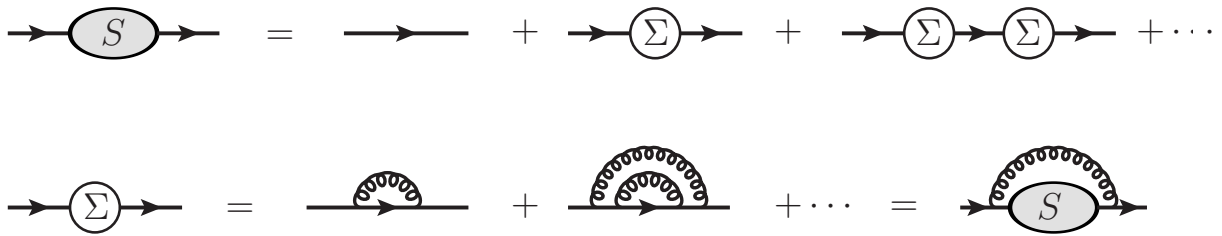}
 	\caption{Dyson-Schwinger equation for the dressed quark propagator in large $N$ limit.
 		\label{DS:equation:dressed:quarks}}
 \end{figure}
 %-------------------------

Let $S(p^\mu)$ denote the full (dressed) quark propagator, $\Sigma(p^\mu)$ signify the 1PI quark self-energy.
In the large $N$ (planarity) limit, the standard rainbow approximation in the Dyson-Schwinger equation for the quark self-energy,  as pictorially depicted in Fig.~\ref{DS:equation:dressed:quarks}, becomes a rigorous procedure:
%-------------------
\begin{subequations}
%-------------------
\bqa
%-------------------
&& S(p^\mu)={i\over \not\! p-m-\Sigma(p^\mu)+i\epsilon},
%-------------------
\label{dressed:quark:propagator}
%-------------------
\\
%-------------------
&& \Sigma(p^\mu) =
{\lambda\over 2\pi}  \pvint{-\infty}{\infty} {dk^0 dk^1\over (k^1-p^1)^2} \gamma^0 S(k^\mu)\gamma^0,
%-------------------
 \label{DS:dressed:quark:1PI:self:energy}
%-------------------
%-------------------
\eqa
%-------------------
 \label{Dyson:Schwinger:eq:quark:propagator}
%-------------------
\end{subequations}
%-------------------
where $\pvint{}{}$ implies a principal-value prescription.

It turns out that the quark self-energy $\Sigma$ only depends on the spatial component
of the two-vector $p^\mu$, and can be parameterized as
%-------------------
\beq
%-------------------
\Sigma(p^1)= A(p^1) + B(p^1) \gamma^1.
%-------------------
%-------------------
\eeq
%-------------------

For notational brevity, from now on we often use the symbol $p$ to represent $p^1$, unless otherwise explicitly stated.
It is convenient to introduce two new variables $E(p)$ and $\theta(p)$, in replacement of the
the functions $A(p)$ and $B(p)$:
%-------------------
\begin{subequations}
%-------------------
\bqa
%-------------------
&& A(p)= E(p)\cos\theta(p)-m,
%-------------------
\\
%-------------------
&& B(p)= E(p)\sin \theta(p)-p,
%-------------------
%-------------------
%-------------------
\eqa
%-------------------
 \label{Intro:Ep:thetap}
%-------------------
\end{subequations}
%-------------------
where $E(p)$ characterizes the energy dispersion of the dressed quark, and $\theta(p)$ is usually referred to as the Bogoliubov (chiral) angle.

Integrating  (\ref{DS:dressed:quark:1PI:self:energy}) over $k^0$, and after some algebra, one arrives at the so-called mass-gap equation~\cite{Bars:1977ud}:
%-------------------
\beq
%-------------------
p \cos \theta(p) -m \sin\theta(p)= \frac{\lambda}{2} \pvint{-\infty}{\infty} \frac{dk}{(p-k)^2} \sin(\theta(p)-\theta(k)).
%-------------------
\label{eq:mass:gap}
%-------------------
\eeq
%-------------------
$\theta(p)$ is an odd, and, monotonically rising function in $p$, which approaches $\pm {\pi\over 2}$ as $p\to \pm \infty$,
respectively.
It turns out to be an impossible mission to express $\theta(p)$ in terms of the known special functions.
As matter of fact, this nonlinear integral equation can only be solved numerically, even in the chiral limit.
Notice in the free theory ($\lambda=0$) limit, $\theta(p)=\tan^{-1}(p/m)$,
recovers the familiar Foldy-Wouthuysen angle in the free Dirac theory.

Interestingly, Eq.~(\ref{eq:mass:gap}) can also be derived from an alternative perspective, {\it viz}, by the requirement of
minimizing the vacuum energy~\footnote{We remind the readers that, in the original BG paper~\cite{Bars:1977ud},
a naive step function $\theta(p) = {\pi\over 2}\epsilon(p)$ (where the sign function $\epsilon(p)$ equals 1 when $p>0$,
equals $-1$ when $p<0$) is advocated as the solution of the gap equation in
the chiral limit. It turns out that this ansatz of $\theta(p)$ leads to a
chiral symmetric vacuum state, which bears an infinite higher energy than the true ground state, thus unacceptable~\cite{Kalashnikova:2001df}.}.

Once the $\theta(p)$ is determined, one can proceed to determine the dispersive law of the dressed quark~\cite{Bars:1977ud}:
%-------------------
\beq
%-------------------
E(p)= m \cos \theta(p) + p \sin \theta(p) + \frac{\lambda}{2} \pvint{-\infty}{\infty} \frac{dk}{(p-k)^2} \cos(\theta(p)-\theta(k)),
%-------------------
\label{dispersion:law:dressed:quark}
%-------------------
\eeq
%-------------------
which is clearly an even function of $p$. Note this dispersive relation is not even Lorentz covariant.
This can be attributed to the fact that, since the Poincar\'{e} algebra does not close in colored sector,
so the Lorentz covariance is scarified in a single quark sector, though it must hold in color-singlet channel.

%-------------------------
 \begin{figure}[htbp]
 	\centering
 	\includegraphics[width=\textwidth]{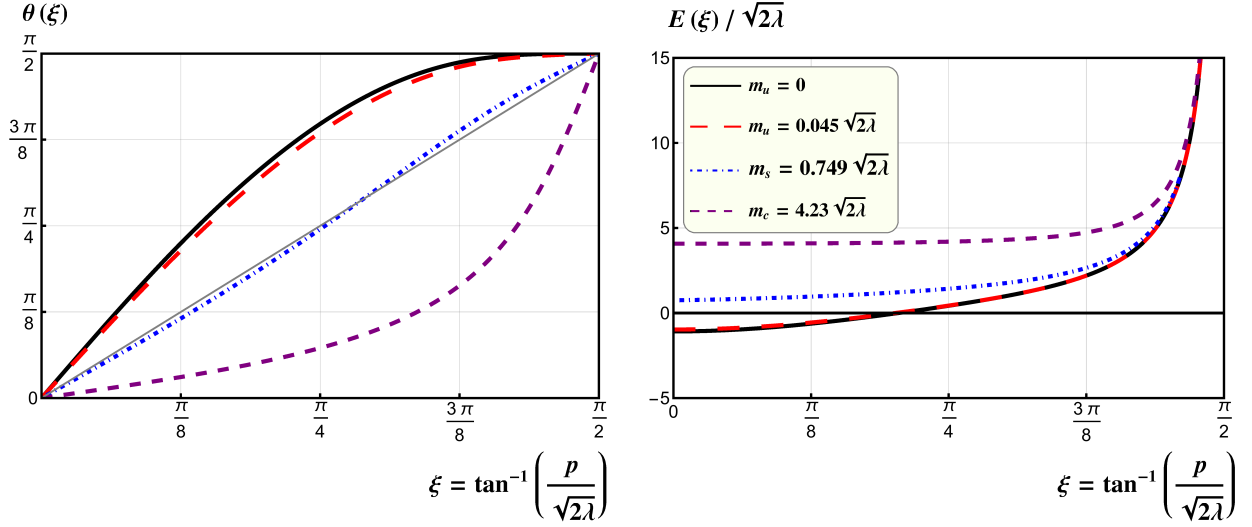}
 	\caption{Bogoliubov (chiral) angle $\theta(p)$ and dressed quark energy $E(p)$ as functions of
 $\xi=\tan^{-1} {p\over \sqrt{2\lambda}}$
 for different current quark mass. \label{theta_omega}}
 \end{figure}
 %-------------------------

As will be elaborated in Section~\ref{Numerical:recipes}, we solve the mass gap equation (\ref{eq:mass:gap}) numerically
using Newton method.
In Fig.~\ref{theta_omega}, we plot the profiles of the chiral angle and dispersive law as function of
quark momentum, for several different quark mass. We see that the Bogoliubov angle in chiral limit indeed
assumes a nontrivial shape.
For small bare quark mass, when the quark momentum is very soft, the dressed quark energy may even become negative.
This pathological behavior can be readily seen from
Fig.~\ref{theta_omega}, which can also be understood from the approximate formula
$E(0)\approx m-{\pi \lambda\over 8}\theta'(0)$~\cite{Kalashnikova:2001df}.

\subsection{The Bars-Green Equation}

\begin{figure}[htbp]
 	\centering
 	\includegraphics[width=0.75\textwidth]{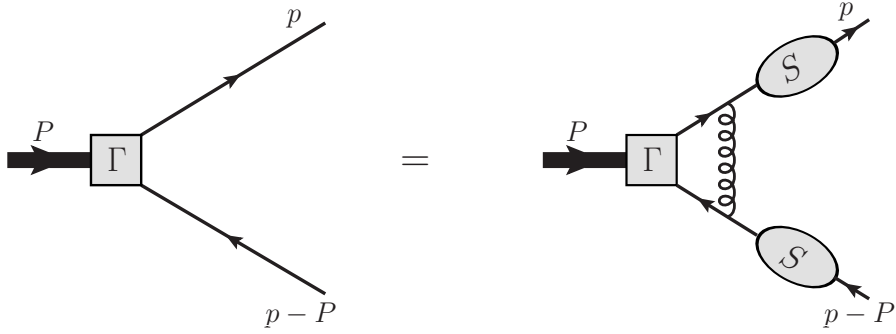}
 	\caption{Bethe-Salpeter equation for the meson-quark-antiquark vertex.
 		\label{Fig:Bethe:Salpeter:eq}}
 \end{figure}
 %-------------------------

As a confining theory, ${\rm QCD}_2$ admits an infinite tower of stable color-singlet mesons in the large $N$ limit.
We are interested in inferring the bound-state equation from the familiar Bethe-Salpeter approach, although the alternative approach, {\it i.e.}, the Hamiltionian method, may be particulary illuminating in certain aspects.
 For simplicity, throughout this work we have
focused on the flavor-neutral quarkonium state, though the extension to the flavored mesons
are straightforward.

The meson-quark-antiquark vertex, denoted by $\Gamma(p^\mu,P^\mu)$,
obeys the homogenous Bethe-Salpeter equation:
%-------------------
\beq
%-------------------
\Gamma(p^\mu,P^\mu)= {i \lambda\over 2\pi} S(p^\mu) \pvint{-\infty}{\infty}
{d k^0 d k \over (p-k)^2} \gamma^0 \Gamma(k^\mu,P^\mu)\gamma^0 S(p^\mu-P^\mu),
%-------------------
\label{Bethe-Salpeter:equation:axial:gauge}
%-------------------
\eeq
%-------------------
where $P^\mu$ is the two-momentum of the meson, and $p^\mu$($P^\mu-p^\mu$) is the momentum of the external quark(antiquark) leg.
This equation is pictorially represented in Fig.~\ref{Fig:Bethe:Salpeter:eq}, where the ladder approximation
also becomes justified, thanks to the planarity condition.

It is a standard practice to introduce the Bethe-Salpeter wave function:
%-------------------
\beq
%-------------------
\Phi(p,P^\mu) \equiv \int d p^0 \Gamma(p^\mu,P^\mu),
%-------------------
%-------------------
\eeq
%-------------------

Before proceeding, it is convenient to rewrite the full quark propagator in (\ref{dressed:quark:propagator}) as
%-------------------
\begin{subequations}
%-------------------
\bqa
%-------------------
&& S(p^\mu) = {\Lambda_+(p)\gamma^0 \over p^0-E(p)+i\epsilon}+{\Lambda_-(p)\gamma^0
\over p^0+E(p)-i\epsilon},
%-------------------
\\
%-------------------
&& \Lambda_\pm(p) = T(p) {1\pm \gamma^0\over 2} T^\dagger(p),
%-------------------
\\
%-------------------
&&  T(p) = e^{-{1\over 2}\theta(p)\gamma^1}.
%-------------------
\eqa
%-------------------
 \label{Quark:propagator:rewrite}
%-------------------
\end{subequations}
%-------------------
It is also useful to decompose the Bethe-Salpeter matrix wave function $\Phi$
into a pair of wave functions $\phi_\pm$:
%-------------------
\beq
%-------------------
\Phi(p,P^\mu) =  T(p) \left( {1+\gamma^0 \over 2}\gamma^5  \phi_+(p,P)+{1-\gamma^0 \over 2}\gamma^5 \phi_-(p,P)\right) T^\dagger(P-p).
%-------------------
%-------------------
\eeq
%-------------------

Substituting (\ref{Quark:propagator:rewrite}) into (\ref{Bethe-Salpeter:equation:axial:gauge}),
and integrating both sides over $p^0$ by employing the method of residues,
one ends up with two coupled equations for each mesonic state with mass $M_n$ (of the $n$~th mesonic level)~\cite{Bars:1977ud}:
%-------------------
\begin{subequations}
%-------------------
\bqa
%-------------------
&& (E(p)+E(P-p)-P^0) \phi_+(p,P)
= \lambda \pvint{-\infty}{\infty} {dk \over (p-k)^2}
\Big[C(p,k,P) \phi_+(k,P) - S(p,k,P)  \phi_-(k,P)\Big],
%-------------------
\nn\\\\
%-------------------
&&  (E(p)+E(P-p)+P^0)  \phi_-(p,P)
= \lambda \pvint{-\infty}{\infty} {dk \over (p-k)^2}
\Big[C(p,k,P) \phi_-(k,P) - S(p,k,P)  \phi_+(k,P)\Big],
%-------------------
\nn\\
%-------------------
%-------------------
\eqa
%-------------------
\label{Bars:Green:equations:for:mesons}
%-------------------
\end{subequations}
%-------------------
where $P^\mu P_\mu = M_n^2$, the dressed quark energy
$E(p)$ is given in (\ref{dispersion:law:dressed:quark}), and
%-------------------
\begin{subequations}
%-------------------
\bqa
%-------------------
&& C(p,k,P) =
\cos {\theta(p)-\theta(k)\over 2} \cos {\theta(P-p)-\theta(P-k)\over 2},
%-------------------
\\
%-------------------
&&  S(p,k,P) =
\sin {\theta(p)-\theta(k)\over 2} \sin {\theta(P-p)-\theta(P-k)\over 2},
%-------------------
\eqa
%-------------------
\label{Definition:C:S}
%-------------------
\end{subequations}
%-------------------
where the Bogoliubov angle $\theta(p)$ is deduced from solving the gap equation (\ref{eq:mass:gap}).

Eq.~(\ref{Bars:Green:equations:for:mesons}) is the mesonic bound-state equation in axial gauge
with equal-time quantization, hereafter referred to as the Bars-Green (BG) equation.
It is the instant-form counterpart of the 't Hooft equation, (\ref{tHooft:equation}).
Unlike a single meson wave function $\phi$ in light-front formalism, here one must introduce a pair of meson wave functions
$\phi_\pm$, in order to warrant the Lorentz covariance in the equal-time quantization.
The much more sophisticated form of the BG equation with respect to the 't Hooft equation, simply reflects the
widely-spread tenet, that boosting the equal-time bound-state wave function
is a highly nontrivial mission, in sharp contrast to the boost-invariant LF formulation.

The wave functions $\phi_\pm$, representing the large (small) component of bound-state solution,
respectively,
characterize the probability amplitude for the $q\bar{q}$ pair moving forward (backward) in time.
Their physical meanings become even more transparent in the bosonization approach,
where the $\phi_\pm$ can be directly interpreted as the coefficient functions associated with a
Bogoliubov transformation from the composite quark-antiquark creation operator to the
mesonic creation operator~\cite{Kalashnikova:2001df}.
Specifically speaking, there are two ways to produce a mesonic state from the vacuum
in the equal-time quantization.
One can always produce a meson state by creating a $q\bar{q}$ pair out of the vacuum,
regardless of the (non)trivial nature of the vacuum, with the probability amplitude characterized by $\phi_+$.
Nevertheless, when the vacuum is nontrivial, {\it e.g.},
which may accommodate a nonzero quark condensate, as what is happening in the equal-time formulation for
large $N$ ${\rm QCD}_2$, one can also create a meson by {\it removing} a redundant pair of $q\bar{q}$ from some correlated quark-antiquark pairs constantly fluctuating out of the nontrivial vacuum.
It is intuitively conceivable that, the relative magnitude of $\phi_-$ with respect to
$\phi_+$ gets more and more suppressed with the increasing quark mass/meson momentum/principal quantum number.

The meson wave functions $\phi_\pm$ obey the following orthogonality and completeness conditions~\cite{Kalashnikova:2001df}~\footnote{In early works,
the orthogonality conditions for $\phi_\pm$ seem to be prescribed in an {\it ad hoc} manner.
For example, in Ref.~\cite{Li:1987hx}, the normalization condition
is such that a {\it positive} sign is chosen between $(\phi_+)^2$ and $(\phi_-)^2$ in the integrand.}:
%-------------------
\begin{subequations}
%-------------------
\bqa
%-------------------
&& \int_{-\infty}^\infty dp \left(\phi_+^{(n)}(p,P) \phi_+^{(m)}(p,P) - \phi_-^{(n)}(p,P) \phi_-^{(m)}(p, P)\right) = \left|P\right| \,\delta^{mn},
%-------------------
\label{Normalization:cond:BG:most:important}\\
%-------------------
&& \int_{-\infty}^\infty dp \left(\phi_+^{(n)}(p,P) \phi_-^{(m)}(p,P) - \phi_-^{(n)}(p,P) \phi_+^{(m)}(p,P)\right) = 0,
%-------------------
%-------------------
\\
%-------------------
&& \sum_{n=0}^\infty \left(\phi_+^{(n)}(p,P) \phi_+^{(n)}(k, P) - \phi_-^{(n)}(p,P) \phi_-^{(n)}(k,P) \right) = \left|P\right|\,\delta(p-k).
%-------------------
\\
%-------------------
&&\sum_{n=0}^\infty \left(\phi_+^{(n)}(p,P) \phi_-^{(n)}(k,P) - \phi_-^{(n)}(p,P) \phi_+^{(n)}(k,P) \right)  = 0.
%-------------------
%-------------------
\eqa
%-------------------
\label{Normalization:condition}
%-------------------
\end{subequations}
%-------------------
The ubiquitous minus signs are reminiscent of the nature of Bogoliubov transformation,
which are manifest using the Hamilton approach~\footnote{Note that $\phi_\pm$ in (\ref{Normalization:condition}) differ
from what are given in Ref.~\cite{Kalashnikova:2001df} by a factor $\sqrt{\frac{\left|P\right|}{2\pi}}$, with the advantage
that our $\phi_+$ in the IMF exactly reduces to the 't Hooft wave function $\phi(x)$, which is cannonically normalized
as $\int^1_0\,dx\, |\phi(x)|^2=1$.}.

It is instructive to examine the properties of the wave functions under discrete symmetry transformation.
Since ${\rm QCD}_2$ is symmetric under space inversion, charge conjugation, the flavor-neutral quarkonium
wave functions must be subject to the following relations:
%-------------------
\begin{subequations}
%-------------------
\bqa
%-------------------
&& \phi_{\pm}^{(n)}(-p,-P) =  -\eta_n \phi_{\pm}^{(n)}(p,P)\qquad\qquad\;{\texttt{P}},
%-------------------
\label{Space:inversion:transformation}
%-------------------
\\
%-------------------
&& \phi_{\pm}^{(n)}(P-p, P) = (-)^n \phi_{\pm}^n(p,P) \qquad\qquad{\texttt{C}},
%-------------------
\label{charge:conjugation:transformation}\\
%-------------------
&& \phi_{\pm}^{(n)}(p-P,-P)= \phi_{\pm}^{(n)}(p,P) \qquad\qquad\quad {\texttt{CP}},
%-------------------
\eqa
%-------------------
\label{Wave:functions:parity:properties}
%-------------------
\end{subequations}
%-------------------
where $\eta_n= (-)^{n+1}$ signals the intrinsic parity of each meson~\footnote{Since in this work
we are only considering a single flavor, each meson also bears a
$C$-parity $\eta^C_n= (-)^n$.}.
Thus all the mesonic levels simply alternate in parity: parity-odd states ($n=0,2,4,\ldots$)
and parity-even states ($n=1,3,5,\ldots$). These symmetry relations signal
a notable virtue of Bars-Green formalism versus LF formulation, since it
is far from straightforward to realize the parity transformation in the latter setup.

A very special case is the ground-state meson in the chiral limit, which turns out to be
parity-odd and exactly massless. For this reason, it is often dubbed {\it chiral pion}, $\pi_\chi$.
Its BG wave functions are known in a semi-analytical fashion~\cite{Kalashnikova:2001df}:
%-------------------
\beq
%-------------------
 \phi_\pm^{\pi_\chi}(p,P) = {1\over 2}\left(\cos{\theta(P-p)-\theta(p)\over 2}
 \pm \sin{\theta(P-p)+\theta(p)\over 2}\right),\qquad{\text{for } P>0},
%-------------------
\label{BG:wave:func:chiral:pion:P:positive}
%-------------------
\eeq
%-------------------
where $\theta(p)$ is the corresponding Bogoliubov angle in the chiral limit.

Applying (\ref{Space:inversion:transformation}) to (\ref{BG:wave:func:chiral:pion:P:positive}),
we can obtain the BG wave functions of the chiral pion that moves toward the negative $x$-axis:
%-------------------
\beq
%-------------------
 \phi_\pm^{\pi_\chi}(p,P) = {1\over 2}\left(\cos{\theta(|P|+p)+\theta(p)\over 2}
 \pm \sin{\theta(|P|+p)-\theta(p)\over 2}\right),\qquad{\text{for } P<0}.
%-------------------
\label{BG:wave:func:chiral:pion:P:negative}
%-------------------
\eeq
%-------------------
Since one cannot boost a massless particle to its rest frame,
some sort of irregularity is anticipated to occur in the $P\to 0$ limit.
For a fixed $p$, the BG wave functions for a chiral pion,
$\phi_\pm^{\pi_\chi}(p,P)$, are continuous yet nonanalytic across the point $P=0$,
as indicated in
(\ref{BG:wave:func:chiral:pion:P:positive}) and (\ref{BG:wave:func:chiral:pion:P:negative}).

\subsection{Bars-Green equation in IMF}
\label{BG:equation:boosted:to:IMF}

Examining the the coupled integral equations
(\ref{Bars:Green:equations:for:mesons}),  it is by no means transparent to prove the
meson mass spectra are frame-independent.
Nevertheless, Bars and Green have argued that, in the IMF,
the backward-moving wave function $\phi_-$ must diminish, so the BG functions must reduce to the celebrated 't Hooft equation,
consequently the forward-moving wave function $\phi_+$ can be identified with the 't Hooft wave function $\phi(x)$.
Bars and Green have already outlined all the necessary clues for the proof.
Nevertheless, for the sake of completeness and clarity, we decide to supplement more technical details in
intermediate steps, together with some pictorial evidences,
to validate Bars and Green's claim.

Let a flavor-neutral meson carry the nonzero momentum $P>0$.
Let us first introduce a pair of dimensionless ratios $x,y$, by $x =p/P$ and $y=k/P$, where $p$ and $k$ represent
the quark momenta appearing in (\ref{Bars:Green:equations:for:mesons}).
We subsequently reexpress the Bars-Green wave function as ${\phi}_\pm(x,P)\equiv \phi_{\pm}(p=xP, P)$.
At this stage, the range of $x$ and $y$  remains unbounded.
Let us temporarily assume $x$, $y$ are not in proximity to 0.
In the IMF limit $P\to \infty$, the Bogoliubov angle $\theta(x P)$ is then approaching its asymptotic values,
${\pi\over 2}\epsilon(x)$. Consequently, one finds in the $P\to +\infty$ limit,
%-------------------
\begin{subequations}
%-------------------
\bqa
%-------------------
 C(x,y,P)\longrightarrow & & \cos \left[ {\pi\over 4}(\epsilon(x)-\epsilon(y))\right]
\cos\left[{\pi\over 4}(\epsilon(1-x)-\epsilon(1-y))\right]
%-------------------
\nn \\
%-------------------
& & =\Theta(x y)\Theta((1-x)(1-y)),
%-------------------
\\
%-------------------
 S(x,y,P)\longrightarrow & & \sin \left[{\pi\over 4}(\epsilon(x)-\epsilon(y))\right]
 \sin\left[{\pi\over 4}(\epsilon(1-x)-\epsilon(1-y))\right]
%-------------------
 \nn\\
%-------------------
& &= -\Theta(-x y)\Theta(-(1-x)(1-y)),
%-------------------
\eqa
%-------------------
\label{IMF:limit:C:S}
%-------------------
\end{subequations}
%-------------------
where $\Theta$ designates the Heaviside step function. Therefore, the $C$ function
equals 1 if $0<x,y<1$, or $x, y <0$, or $x, y>1$, and vanishes in all other cases;
The $S$ function always vanishes except when $x<0, y>1$ or $x>1, y<0$, in which cases
it equals -1.

%-------------------------
 \begin{figure}[htbp]
 	\centering
 	\includegraphics[width=\textwidth]{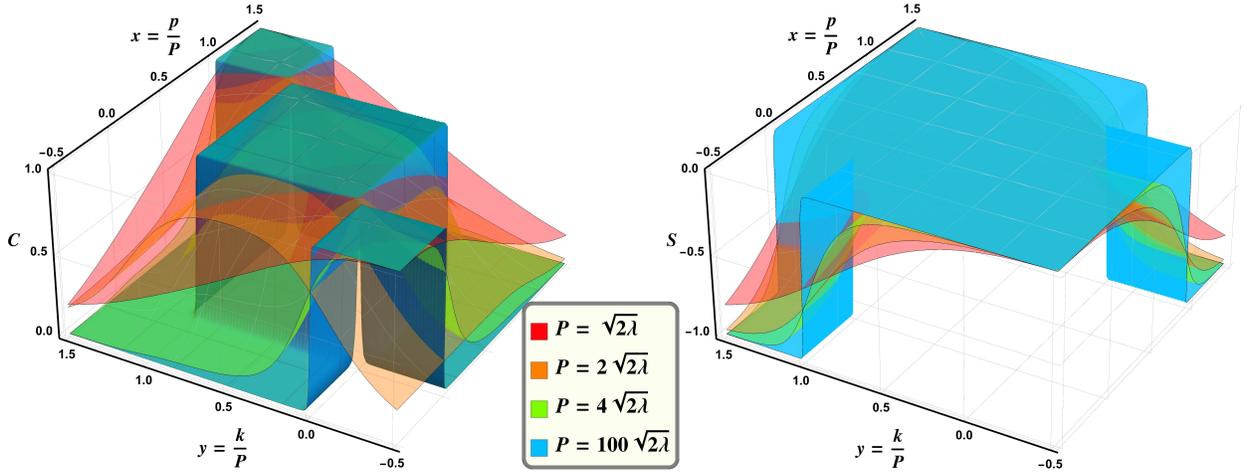}
 	\caption{The $C$ and $S$ functions viewed in various reference frames.
 As a concrete example, here we generate the Bogoliubov angle for the massless $u$ quark.
\label{Fig:C:S:function:different:meson:momentum}}
 \end{figure}
 %-------------------------

For the sake of clarity, we take the chiral limit as a concrete example, with the respective Bogoliubov angle $\theta(p)$
shown in Fig.~\ref{theta_omega}. We then generate the corresponding $C(x,y,P)$ and $S(x,y,P)$ functions,
with several different choice of $P$. From Fig.~\ref{Fig:C:S:function:different:meson:momentum},
we clearly see the trend that, in the IMF limit,
$C(x,y,P)$ and $S(x,y,P)$ indeed exhibit the behavior as dictated in (\ref{IMF:limit:C:S}).

Following Ref.~\cite{Bars:1977ud},
it is straightforward to see that the right-hand sides of Bars-Green equations (\ref{Bars:Green:equations:for:mesons}) must scale as ${\cal O}(1/P)$ in the IMF limit. Therefore, any term in the left-hand side
which scales as $P^1$ or $P^0$ must cancel, and the  ${\cal O}(1/P)$ terms in both sides must be matched.
By examining the asymptotic behavior for the factor $E(p)+E(P-p)\pm P^0$,
we thereby find that $\phi_-(x,P)$ must vanish for all $x$, and
$\phi_+(x)$ is non-vanishing only when $0\le x \le 1$.
Notice than, for $0\le x \le 1$, Eq.~(\ref{IMF:limit:C:S}) then implies that
$C(x,y,P\to \infty)\to \Theta(y(1-y))$, and, $S(x,y,P\to \infty)\to 0$.

From the gap equation (\ref{eq:mass:gap}), it is easy to see that
$\tan\theta(xP)={x P\over m}+{\cal O}(1/P)$ in the $P\to \infty$ limit.
Therefore, the dispersive law in
(\ref{dispersion:law:dressed:quark}) in the IMF simplifies into
%-------------------
\bqa
%-------------------
 E(x P)&\longrightarrow & |x| P + {m^2\over |2x| P}+{\lambda\over 2P}\pvint{-\infty}{\infty} {dy \over (x-y)^2}
\cos{\pi\over 2}(\epsilon(x)-\epsilon(y))+ {\cal O}\left({1\over P^2}\right)
%-------------------
\nn\\
%-------------------
&=&  |x| P + {m^2-2\lambda\over |x| P}+{\cal O}({1/P^2}).
%-------------------
%-------------------
\eqa
%-------------------

Employing the aforementioned simplifications in the IMF,
the Bars-Green equation (\ref{Bars:Green:equations:for:mesons}) then reduces to
%-------------------
\beq
%-------------------
\left({m^2-2\lambda\over x(1-x)P}+ P - P^0\right)\phi_+(x,P)= {\lambda\over P} \pvint{0}{1} {dy\over(x-y)^2}
\phi_{+}(y,P).
%-------------------
\label{BG:equation:IMF}
%-------------------
\eeq
%-------------------
Approximating $P^0=\sqrt{P^2+M^2}$ by $P+M^2/2P$, and matching both sides of (\ref{BG:equation:IMF})
through the linear order in $1/P$, one finds that
%-------------------
\begin{subequations}
%-------------------
\bqa
%-------------------
&& \left(\frac{m^2-2\lambda}{x\left(1-x\right)}-M^2\right)
\phi_+(x,P)=2\lambda\pvint{0}{1}dy
\frac{\phi_+\left(y,P\right)}{(x-y)^2},
%-------------------
\\
%-------------------
&& \phi_-(x,P)={\mathcal O}(1/P^2).
%-------------------
\eqa
%-------------------
\label{IMF:limit:Bars-Green:equation}
%-------------------
\end{subequations}
%-------------------
As promised, the BG equation (\ref{Bars:Green:equations:for:mesons}) for $\phi_+(x,P)$
in IMF does reduce to the 't Hooft equation (\ref{tHooft:equation}) (with $m_1=m_2=m$),
while $\phi_-(x, P)$ dies away with a pace $\propto {1\over P^2}$.
In the following sections, we will numerically examine the tendency of the BG wave functions $\phi_\pm$
with an ever-increasing meson momentum.

\section{Some Lorentz-invariant quantities in axial gauge}
\label{Chiral:condensate:decay:constants}

There are some basic yet important nonperturative quantities, exemplified by the quark vacuum condensate (for arbitrary quark mass) and meson decay constants, which have been extensively studied in the LF formulation of ${\rm QCD}_2$.
In this section, we revisit these quantities in the axial gauge in equal-time quantization.
To our knowledge, the studies from the perspective of Bars-Green formalism are novel.
The purpose of this Section is to make a nontrivial examination of the gauge and Lorentz invariance
(frame independence) of these simple QCD matrix elements.

\subsection{Quark condensate}

Since the mid-80s, it became widely known that the $1+1$-dimensional QCD in the large $N$ limit
actually accommodates spontaneous chiral symmetry breaking (SCSB),
signalled by the non-zero quark condensate~\cite{Zhitnitsky:1985um,Li:1986gf}.

In passing, it is worthwhile to elaborate on the possible phases in the 't Hooft model.
The massless ${\rm QCD}_2$ ($N\to \infty$) can be classified in two
distinct regimes, depending on the order of taking the $N\to\infty$ and the chiral limit,
which turn out not to commute~\cite{Zhitnitsky:1985um}:
$1)$ In the weak coupling regime, one assumes $m_q \gg g \sim {1\over \sqrt{N}}$, and the $N\to\infty$ limit is
taken prior to ultimately sending $m_q\to 0$. This phase corresponds to the familiar mass spectrum from
solving 't Hooft equation, where the spontaneous chiral symmetry breaking occurs.
$2)$ In the strong coupling regime, where $m\ll g \sim {1\over \sqrt{N}}$ is instead assumed, and one first
takes the chiral limit, then followed by sending $N\to\infty$.
Chiral symmetry remains unbroken in this phase, and the corresponding spectrum is rather different,
where there appear massless composite fermion rather than the massless meson~\cite{Baluni:1980bw,Steinhardt:1980ry,Amati:1981fn,Bhattacharya:1981gy,Berruto:1999cy}.
In this work, we have tacitly assumed to exclusively consider the weak-coupling phase.

At first sight, the occurrence of SCSB in ${\rm QCD}_2$ (with $N\to\infty$)
appears to contradict Coleman's theorem~\cite{Coleman:1973ci},
which seems to rule out the possibility of spontaneous breakdown of any continuous symmetry
in two dimensional field theory. This puzzle was first resolved by Witten~\cite{Witten:1978qu}
in the context of the $SU(N)$ Thirring model in the $N\to\infty$ limit
(this model is also commonly referred to as the Gross-Neveu model).
He pointed out that the Berezinskii-Kosterlitz-Thouless (BKT) phenomenon actually occurs in this case~\cite{Berezinskii:1973,Kosterlitz:1973xp}, so that the chiral symmetry is
``almost'' spontaneously broken. Later Zhitnitsky realized that, in the weak coupling regime,
the 't Hooft Model also exhibits exactly the same BKT effect~\cite{Zhitnitsky:1985um},
so that the SCSB also occurs in the $N\to\infty$ limit.
The spontaneous chiral symmetry breaking is consistent with the spectrum of 't Hooft model that
the mesonic states with opposite $P$-parities are non-degenerate in the masses.

Specifically speaking, one can show that the following two-point correlator
in the `t Hooft model possesses the following large-$|x|$ behavior~\cite{Zhitnitsky:1985um}:
%-------------------
\beq
%-------------------
\langle 0\vert \bar{q}(x)(1+\gamma_5)q(x) \bar{q}(0)(1-\gamma_5)q(0)
\vert 0 \rangle \sim |x|^{-1/N}.
%-------------------
\label{correlator:large:x}
\eeq
%-------------------
In the $N\to\infty$ limit, the correlator approach a non-vanishing constant, thus
exhibiting the true long-range order, and heralding the occurrence of the massless boson mode;
for any large but finite $N$, the correlator falls off very slowly with $x$,
and there does not arise massless meson.
Hence there is no contradiction with Coleman's theorem.

Despite the notion of pertubative vacuum in the light-front quantization,
a nonvanishing chiral condensate was first discovered from this formalism.
Using the operator expansion technique, Zhitnistsky has found an analytic result for the vacuum quark condensate
in the chiral limit in 't Hooft model~\cite{Zhitnitsky:1985um}:
%-------------------
\beq
%-------------------
\langle\bar{q} q \rangle \Big|_{m=0} = - {N\over \sqrt{6}}\sqrt{\lambda}.
%-------------------
\label{chiral:condensate:chiral:limit:Zhitnistsky}
\eeq
%-------------------
Since the condensate is nonanalytic in the 't Hooft coupling $\lambda$, it characterizes
a type of nonperturbative effect that cannot be captured by summing perturbation series in $\lambda^n$ ($n$ being a non-negative integer) to all orders.

Later, Burkardt has extended Zhinitstsky's analysis, and presented an analytic formula also for
massive quark~\cite{Burkardt:1995eb}.
After subtracting the logarithmic UV divergence, he obtains the renormalized
quark condensate for an arbitrary value of $m$:
%-------------------
\beq
%-------------------
\langle\bar{q} q \rangle_{\rm ren} = {N m\over 2\pi} \left\{\ln (\pi\alpha)-1-\gamma_E+
\left(1-{1\over \alpha}\right) \left[(1-\alpha)I(\alpha)-\ln 4\right] \right\},
%-------------------
\label{condensate:finite:m:LF}
\eeq
%-------------------
where $\alpha=2\lambda/m^2$, $\gamma_E=0.5772\ldots$ is the Euler constant, and
%-------------------
\beq
%-------------------
I(\alpha)= \int^\infty_0 {dy\over y^2} {1-{y\over \sinh y \cosh y}\over 1+\alpha(y \coth y -1)}.
%-------------------
\eeq
%-------------------

A nontrivial vacuum state naturally emerges in ${\rm QCD}_2$ if equal-time quantization is taken.
It was first by Li~\cite{Li:1986gf} who first reported a nonzero quark condensate
in the chiral limit in the axial gauge:
%-------------------
\beq
%-------------------
\langle\bar{q} q \rangle \Big|_{m=0} = N\int{dp \over 2\pi}{\rm Tr}\left[\gamma^0\Lambda_-(p)\right]=
-{N\over\pi }\int_0^{\infty} dp \,\cos \theta(p).
%-------------------
\label{chiral:condensate:equal:time:formalism:zero:mass}
\eeq
%-------------------
Substituting the numerical solution of $\theta(p)$ from (\ref{eq:mass:gap}), into this equation,
one readily verifies (\ref{chiral:condensate:chiral:limit:Zhitnistsky}) obtained from light-front formalism, to a high numerical
accuracy.

For nonzero quark mass, the integral in (\ref{chiral:condensate:equal:time:formalism:zero:mass}) becomes logarithmically UV
divergent. Subtracting the analogous term arising from cosine of the Foldy-Wouthysen angle of a free quark,
which amounts to performing an additive renormalization, one finds that
the renormalized quark condensate in axial gauge is
%-------------------
\beq
%-------------------
\langle\bar{q} q \rangle_{\rm ren} =
-{N\over\pi }\int_0^{\infty} dp \left[\cos \theta(p)-{m\over \sqrt{m^2+p^2}}\right].
%-------------------
\label{condensate:finite:m:Equal:Time}
\eeq
%-------------------

\begin{figure}
\centering{}
\includegraphics[width=0.65\textwidth]{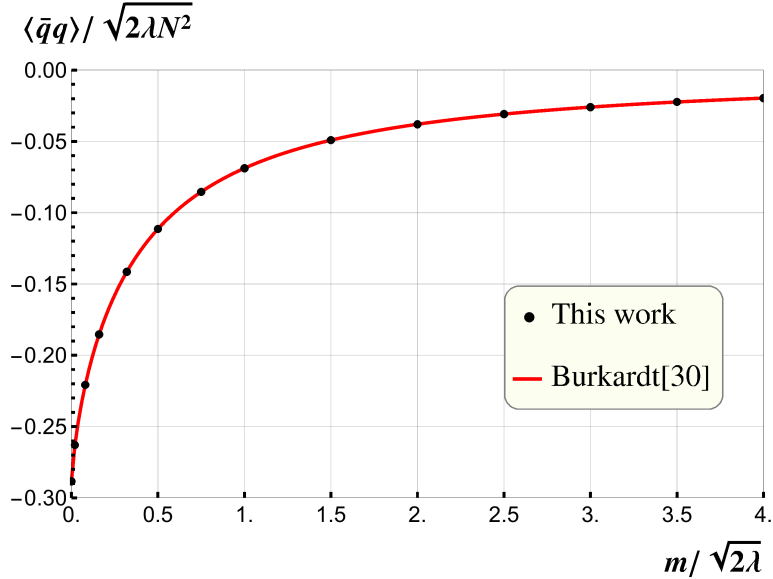}\\
\caption{Renormalized quark condensate as a function of quark mass.\label{fig:Vcm_Cndst}}
\end{figure}

In Fig.~\ref{fig:Vcm_Cndst}, we plot the the quark condensate as function of quark mass, stemming from the
light cone gauge, (\ref{condensate:finite:m:LF}), and the axial gauge, (\ref{condensate:finite:m:Equal:Time}).
Quite satisfactory agreement is achieved, firmly establishing the gauge invariance of the quark condensate.

\subsection{Decay constants}
\label{section:decay:constant}

One can define the meson decay constant $f^{(n)}$  as
%-------------------------
\begin{align}
%-------------------------
& \langle \Omega \left| \bar q\gamma^\mu\gamma^5 q \right|M_n (P) \rangle
%-------------------------
%\\
=
\begin{cases}
%-------------------------
f^{(n)} {P^\mu\over \sqrt{2 P^0}},   \qquad\mbox{$n$\;even}
%-------------------------
\\
%-------------------------
f^{(n)} {\epsilon^{\mu\nu} P_\nu\over \sqrt{2 P^0}}, \qquad\mbox{$n$\;odd},
%-------------------------
%\nn
\end{cases}
%-------------------------
\label{Decay:Constant:definitions}
\end{align}
%-------------------------
%-------------------------
where $\epsilon_{\mu\nu}$ is the antisymmetric Levi-Civita tensor in two dimensions.

In the light-cone gauge, Callan, Gross and Coote~\cite{Callan:1975ps}
were able to identify the decay constant for the $n$-th mesonic level
simply with the integral of the 't Hooft wave function:
%-------------------
\beq
%-------------------
f^{(n)} =  \sqrt{N\over \pi} \int_0^1 dx\, \phi^{(n)}(x). %\qquad\mbox{for\;$\mu=+$}.
%-------------------
\label{decay:constant:LC}
\eeq
%-------------------

For the chiral pion, $\pi_\chi$ (the massless parity-odd state affiliated with $m=0$)~\footnote{Witten emphasized
that it would be a misconception to interpret this massless meson as a (decoupled) Goldstone boson~\cite{Witten:1978qu}.
Nevertheless, for the sake of convenience, and, in conformity with most recent literature,
bearing the ``almost'' spontaneous chiral symmetry breaking and
BKT phenomenon in mind, we will frequently refer this massless meson as chiral pion, or Goldstone boson throughout this work.},
the 't Hooft wave function possesses a peculiar form: $\phi^{\pi_\chi}(x)=\Theta(x(1-x))$,
so the decay constant simply is
%-------------------
\beq
%-------------------
f^{\pi_\chi} = \sqrt{N \over \pi}.
\label{LC:decay:constant:chiral:pion}
%-------------------
\eeq
%-------------------

The decay constant in the axial gauge can be most conveniently worked out following the Hamiltonian
method~\cite{Kalashnikova:2001df}. With the aid of the bosonization technique, one can reexpress
the axial vector current in term of meson's creation and annihilation operators,
and readily ascertain the intended decay constant.

We separately discuss the decay constants of mesons with odd and even parity, as designated in
(\ref{Decay:Constant:definitions}). First we consider the mesonic level with even $n$ (odd parity):
%-------------------------
\begin{align}
%-------------------------
\label{Decay:constant:BG:n:even}
& f^{(n\;{\rm even})} =
%-------------------------
\begin{cases}
%-------------------------
\sqrt{N\over \pi} {1\over \sqrt{P P^0}} \int_{-\infty}^{\infty}\! dk  \cos\frac{\theta(P-k)-\theta(k)}{2}\left[\phi^{(n)}_+(k,P)
-\phi^{(n)}_-(k,P)\right],\qquad \text{for }\mu=0;
%-------------------------
\\
%-------------------------
\sqrt{N\over \pi} {1\over P} \sqrt{P^0\over P} \int_{-\infty}^{\infty} \! dk
\sin\frac{\theta(P-k)+\theta(k)}{2}\left[\phi^{(n)}_+(k,P)
+\phi^{(n)}_-(k,P)\right],\qquad \text{for }\mu=1.
%-------------------------
%\nn
\end{cases}
%-------------------------
\end{align}
%-------------------------
These two expressions for the decay constant are obtained by
utilizing the different axial vector Lorentz index in (\ref{Decay:Constant:definitions}).
Although both analytical expressions superficially differ, they are doomed to be equal by Lorentz invariance.
Furthermore, although these expressions explicitly depend on the meson momentum $P$, the frame-independence of the decay constant
enforces some identities that $\theta(p)$ must obey.
In Section~\ref{Numerical:Results:Collections}, we shall present explicit numerical
evidences for the frame/Lorentz-index independence of the meson decay constants.
%------------
Note it is quite delicate to extract the decay constant for a stationary ($P\to 0$)
meson from (\ref{Decay:constant:BG:n:even}).
%------------
We emphasize that the inclusion of the small component of the BG wave function, $\phi_-$,
is crucial to warrant the frame-independence of the decay constant.

It is interesting to examine the decay constant of the chiral pion in the axial gauge.
Substituting the analytic BG wave functions (\ref{BG:wave:func:chiral:pion:P:positive}) into (\ref{Decay:constant:BG:n:even}),
for a pion carrying arbitrary positive momentum $P$,
we find
%-------------------------
\beq
%-------------------------
f^{\pi_\chi} = \sqrt{N\over \pi} {1\over P} \int_{-\infty}^{\infty} \! dk
\cos\frac{\theta(P-k)-\theta(k)}{2} \sin\frac{\theta(P-k)+\theta(k)}{2}.
%-------------------------
\eeq
%-------------------------
Though far from obvious to see why the integral is exactly equal to $P$,
it has to be so to match the LF result for chiral pion, Eq.~(\ref{LC:decay:constant:chiral:pion}).

Next we turn to the mesonic levels with odd $n$ (even parity).
For flavor-neutral mesons, such states have odd $C$ parity ,
so the corresponding 't Hooft wave functions are odd in exchanging $x$ and $1-x$. As a result,
the decay constants simply vanish in line with the prediction from the light-cone gauge, (\ref{decay:constant:LC}).

Notwithstanding this trivially looking result, it is still instructive to
examine these decay constants from the angle of
axial gauge. The respective decay constants in this case read
%-------------------------
\begin{align}
%-------------------------
\label{Decay:constant:BG:n:odd}
& f^{(n\;{\rm odd})} =
%-------------------------
\begin{cases}
%-------------------------
\sqrt{N\over \pi} {1\over \sqrt{P P^0}}  \int_{-\infty}^{\infty} \! dk
\sin\frac{\theta(P-k)+\theta(k)}{2}\left[\phi_+^{(n)}(k,P)
+\phi_-^{(n)}(k,P)\right] = 0,\qquad\text{for }\mu=0;
%-------------------------
\\
%-------------------------
\sqrt{N\over \pi} {1\over P} \sqrt{P^0\over P} \int_{-\infty}^{\infty} \! dk \cos\frac{\theta(P-k)-\theta(k)}{2}\left[\phi_+^{(n)}(k,P)
-\phi_-^{(n)}(k,P)\right] = 0,\qquad \text{for }\mu=1.
%-------------------------
\end{cases}
%-------------------------
\end{align}
%-------------------------
Again we show the expressions extracted from (\ref{Decay:Constant:definitions})
by utilizing two different axial vector Lorentz indices.
Making use of the fact that $\theta(p)$ is an odd function of $p$,
and the odd $C$-parity of the BG wave functions for the
odd-$n$ states as encoded in (\ref{charge:conjugation:transformation}), one can prove that the
integrals in (\ref{Decay:constant:BG:n:odd}) indeed vanish, for all possible meson momentum.

\section{Numerical recipes for solving bound-state equation}
\label{Numerical:recipes}

Numerically solving 't Hooft equation has gained a mature status, so here we just briefly describe the
numerical strategies adopted in this work. This type of equation is usually solved by the spectrum method,
with the solution presumed to be a linear combination of a set of basis functions.
For massive quark ($m\gg \sqrt{2\lambda}$), it proves convenient to invoke the so-called Multhopp method,
which utilize the trigonometric basis functions~\cite{Brower:1978wm}~\cite{Grinstein:1997xk}.
For light quark ($m \le \sqrt{2\lambda}$), yet it is more advantageous to follow 't Hooft's original method~\cite{'tHooft:1974hx},
that adopts a set of basis functions such as $\Psi_n(x)=A x^{\beta_1} (1-x)^{2-\beta_1}+B x^{\beta_2} (1-x)^{2-\beta_2} + \sum_n C_n \sin(n \pi x)$, in which parameters $\beta_{1,2}$ are determined by the boundary conditions $\pi \beta_{1,2} \cot
(\pi \beta_{1,2}) = 1- m^2/2\lambda$. Empirically, $n\sim\mathcal{O}\left(10^1\right)$ is sufficient
to yield stable first three digits.

Prior to solving the Bars-Green equations, one has to first determine the chiral angle $\theta(p)$ to a decent accuracy.
Here we follow Ref.~\cite{Li:1987hx} to use the generalized Newton method.
It is convenient to first change the variable from $p$ to $\xi$ using $p = \sqrt{2\lambda} \tan \xi$, so $\xi \in (-{\pi\over 2},{\pi\over 2})$, within a finite interval.
The mass gap equation in (\ref{eq:mass:gap}) is then discretized to a set of matrix equations:
%-------------------------
\begin{align}
%-------------------------
\tan (\xi_{k}) \cos\big[\theta(\xi_k)\big] - m\sin\big[\theta(\xi_k)\big]=
\frac{1}{4}\sum_{j=-N+1}^{N-1}\begin{cases}
\frac{\pi}{2N}\sec^{2}(\eta_j)\frac{\sin\left[\theta(\xi_k)-\theta(\eta_j) \right]} {\left[\tan(\xi_k)-\tan (\eta_j) \right]^{2}} & j\neq k
%-------------------------
\\
%-------------------------
0 & j=k\;.
%-------------------------
\end{cases}
%-------------------------
\end{align}
%-------------------------
Suppose $N$ is a prescribed large positive integer, designating the size of a grid.
Both variables $\xi_j, \eta_j= {j \pi\over 2N}$ are evenly partitioned on the grid,
with $j$ being an integer obeying $-N\leq  j \leq N$.
In addition, the boundary conditions $\theta(\pm{\pi\over 2}) =\pm{\pi\over 2}$
must be imposed.
Greater $N$ will generally decrease the discretization errors, nevertheless render the
computation more expensive.
Practically, $N=100$ works well for the case of light flavors, i.e. $\pi$ and $s\bar{s}$ mesons.
A finer grid with $N=300$ or higher is required for heavy mesons, {\it i.e.}, for $c\bar{c}$ quarkonium.
The numerical results of the chiral angle $\theta(p)$ and dispersive function $E(p)$ have been
shown in Fig.~\ref{theta_omega}.

Analogous to the 't Hooft equation, Bars-Green equation can be solved by means of the spectrum method as well.
The major complication is due to the emergence of the additional $\phi_-(p,P)$ component,
so one inevitably confronts coupled integral equations.
In the late 80s, Li {\it et al.} solved the BG equations for a variety of {\it stationary} flavor-neutral mesons, choosing the basis functions as the quantum harmonic oscillator's eigen-functions~\cite{Li:1987hx}.

As described in Section~\ref{BG:equation:boosted:to:IMF}, it is convenient to introduce a momentum fraction variable $x=p/P$
for a flavor-neutral meson, with meson momentum denoted by $P$ and quark momentum represented by $p$.
The Bars-Green wave function is then effectively expressed as ${\phi}_\pm(x,P)\equiv \phi_{\pm}(p=xP, P)$.
In contrast to the light-front momentum fraction $x \in [0,1]$ in the 't Hooft wave function,
the range of $x$ in BG wave functions is completely unbounded.
The normalization condition in (\ref{Normalization:cond:BG:most:important}) for the BG wave functions
can be rewritten as
%-------------------------
\begin{equation}
%-------------------------
\int_{-\infty}^{\infty}dx\,\left\{\left|\phi^{(n)}_+(x,P)\right|^2-\left|\phi^{(n)}_-(x,P)\right|^2\right\}=1.
%-------------------------
\end{equation}
%-------------------------

To apply the spectrum method to a moving meson, we generalize Li {\it et al.}'s orthogonal basis functions as
follows~\footnote{We note that, there have been some attempt to solve
the BG equation numerically for a moving meson~\cite{Kuramoto:1994np}.
However, the $\phi_-$ component has been completely neglected thereof,
consequently the Poincar\'{e} invariance is inevitably scarified.}:
%-------------------------
\beq
%-------------------------
\Psi_m(\alpha, x,P)=\sqrt{{\left|P\right|\alpha\over 2^m m!\sqrt{\pi}}}e^{-{\alpha^2 P^2(1-2x)^2 \over 8}}
H_m\left({\alpha P\over 2}(1-2 x)\right),
%-------------------------
\eeq
%-------------------------
where $H_m$ represents the $m$-th Hermite polynomial and
$\alpha$ is a variational parameter that can be tuned to minimize the mass of the ground state.
%-------------------------
\begin{align}
%-------------------------
\label{Decay:constant:BG:n:odd}
& \phi_\pm(x, P) =
%-------------------------
\begin{cases}
%-------------------------
\sum_{m=0}^{N-1} a_m^\pm \Psi_{2m}(\alpha, x,P) ,\qquad\text{n even};
%-------------------------
\\
%-------------------------
\sum_{m=0}^{N-1} b_m^\pm \Psi_{2m+1}(\alpha, x,P) ,\qquad\text{n odd}.
%-------------------------
\end{cases}
%-------------------------
\end{align}
%-------------------------
Solving the original coupled integral equations are then transformed into the matrix eigenvalue problem.
After diagonalization of the $N\times N$ matrix, one can determine the mass spectra of the
first $N$ same-parity mesonic states from the discrete energy eigenstates,
$M_n=\sqrt{(P_n^0)^2-P^2}$, as well as the corresponding
BG wave functions $\phi^{(n)}_\pm (x,P)$.
Practically, for most cases taking $N\approx 20$ appears to be adequate.

Before concluding this section, we describe the principal-value prescription employed in this work
in solving 't Hooft and Bars-Green equations.
To tame severe infrared divergences, two distinct strategies are implemented:
%-------------------
\begin{subequations}
%-------------------
\bqa
%-------------------
&&\pvint{}{} {dy  \over (x-y)^2} f(y)=
\int {dy\over (x-y)^2}\left[f(y)-f(x)-(y-x){df(x)\over dx } \right],
%-------------------
\\
%-------------------
%-------------------
&& \pvint{}{} {dy\over (x-y)^2} f(y)=
\lim_{\epsilon\rightarrow 0}\left[\int^{x-\epsilon} {dy\over (x-y)^2}f(y)
+\int_{x+\epsilon} {dy\over (x-y)^2} f(y)
-{2f(x)\over \epsilon}\right],
%-------------------
%-------------------
\eqa
%-------------------
\end{subequations}
%-------------------
where $f(y)$ is a smooth test function which is regular at $y=x$.
The first recipe is the subtraction method utilized in \cite{Li:1987hx}, while the second is the
Hadamard regularization for hypersingular integral~\cite{Hadamard:1923}.
In practice, both prescriptions yield stable and convergent results.

\section{Numerical Results}
\label{Numerical:Results:Collections}

Being a super-renormalizable theory, ${\rm QCD}_2$ bears the gauge coupling $g$ with unit mass dimension.
In the large $N$ limit, we set the absolute mass scale following the ansatz in Ref.~\cite{Burkardt:2000uu},
{\it e.g.}, choosing the value of the 't Hooft coupling $\lambda$ such that $\pi\lambda = 0.18\;{\rm GeV}^2$,
in conformity to the value of string tension in the realistic ${\rm QCD}_4$.
For notational brevity, we will express any dimensionful quantity in units of $\sqrt{2\lambda}= 340$ MeV
in the rest of this section.

In the hypothetic 1$+$1-dimensional world, we attempt to mimic realistic mesons in ${\rm QCD}_4$ as much as possible.
In the $\sqrt{2\lambda}$ unit, the masses of physical $\pi$ and $J/\psi$ mesons are $M_\pi=0.41$, and $M_{J/\psi}=9.03$,
respectively. Solving the 't Hooft function for ground state, the corresponding quark mass are found to be
$m_u=0.045$~\footnote{Here we use the symbol $u$ to designate the light $u,d$ flavors.
In this work we only consider the neutral mesons composed of a single flavor,
concerning flavor-mixing is a sub-leading effect in $1/N$ expansion.},
$m_c=4.23$.

We have also intentionally invented a fictitious strange quark with $m_s=0.749$.
As can be seen in Fig.~\ref{theta_omega}, it corresponds to a peculiar threshold $\theta(\xi)$ (with $\xi=\tan^{-1} p$),
through which the profile of $\theta(\xi)$ passes from the convex to concave with the increasing quark mass.
This particular strange quark mass is determined by minimizing the relative distance between the $\theta(\xi)$  and
the straight line $\theta(\xi)=\xi$. The lowest-lying strangeonium state has $M_{s\bar{s}}=2.18$.
Naively speaking, the strange quark with $m_s=0.749$ might be thought of severing a threshold,
below which is called light flavor, and above which is called heavy flavor.

\begin{figure}
\begin{centering}
\includegraphics[width=0.65\textwidth]{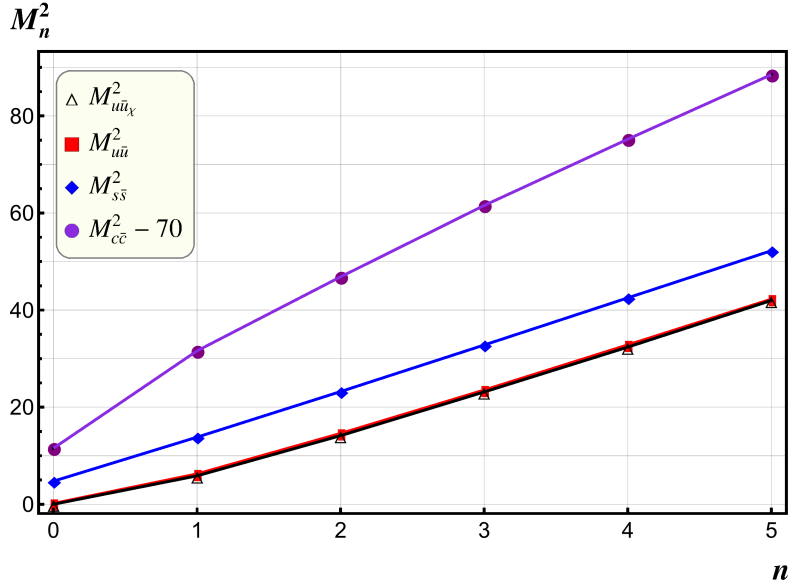}
\par\end{centering}
\caption{Mass spectra of a few low-lying mesonic levels with different quark mass.} \label{fig:Regge}
\end{figure}

For completeness, we also consider the chiral limit with $m_u=0$, which can host a lowest-lying {\it massless} state
named chiral pion. The mass spectra of the first low-lying mesonic levels, for each quark mass,
are listed in Table~\ref{Table:Mass:Spectra} as well as in Fig.~\ref{fig:Regge}. As is well known,
the excited states fit into the linear Regge trajectories to a good degree. We stress that, the mass spectra
found by solving the BG equation appear to be frame-independent, and always agree with what are
obtained by solving 't Hooft equation.
{\it Thus these numerical studies constitute a nontrivial validation of Poincar\'{e} invariance in the
Bars-Green formalism}.

\begin{table}[H]
\begin{centering}
\begin{tabular}{|c|c|c|c|c|c|c|}
\hline
 & $n=0$ & $n=1$ & $n=2$ & $n=3$ & $n=4$ & $n=5$\tabularnewline
\hline
\hline
$M_{u\bar{u}}^{\chi(n)}$ & 0 & 2.43 & 3.76 & 4.81 & 5.68 & 6.46\tabularnewline
\hline
$M_{u\bar{u}}^{(n)}$ & 0.41 & 2.50 & 3.82 & 4.85 & 5.73 & 6.50\tabularnewline
\hline
$M_{s\bar{s}}^{(n)}$ & 2.18 & 3.72 & 4.82 & 5.73 & 6.52 & 7.23\tabularnewline
\hline
$M_{c\bar{c}}^{(n)}$ & 9.03 & 10.08 & 10.81 & 11.47 & 12.05 & 12.59\tabularnewline
\hline
\end{tabular}\caption{Mesonic mass spectra of the $u\bar{u}$ (as well as for the massless $u$ quark), $s\bar{s}$ and $c\bar{c}$ families.
The results are obtained from solving 't Hooft equation, as well as from solving the BG equation in different reference frames.
The respective spectra obtained from different approaches are always compatible with each other,
at least agree at the second decimal digit.
 \label{Table:Mass:Spectra}}
\par\end{centering}
\end{table}

In passing, one might like to take a closer look at the spontaneous chiral symmetry breaking in
${\rm QCD}_2$. The celebrated Gell-Mann-Oakes-Renner relation states that
%---------------------
\beq
%---------------------
M_\pi^2 = -{4 m_u \langle \bar{q} q \rangle\over f_{\pi_\chi}^2}.
%---------------------
\label{GOR:relation}\eeq
%---------------------
With the aid of (\ref{LC:decay:constant:chiral:pion}) and (\ref{chiral:condensate:chiral:limit:Zhitnistsky}),
substituting $f^{\pi_\chi} = 0.564\sqrt{N}$, and $\langle\bar{q} q \rangle \big|_{m=0} = - {N\over \sqrt{12}} =
-0.243 N$, as well as $m_u=0.045$, into the right side of Eq.~(\ref{GOR:relation}), we then predict $M_\pi=0.371$ (126 MeV), which is quite close to the input pion mass $M_\pi=0.41$ (139 MeV).
Thus, the pseudo-Goldstone nature of the ``physical'' pion is explicitly validated.

%---------------------
\begin{figure}
%---------------------
\centering
\includegraphics[width=\textwidth]{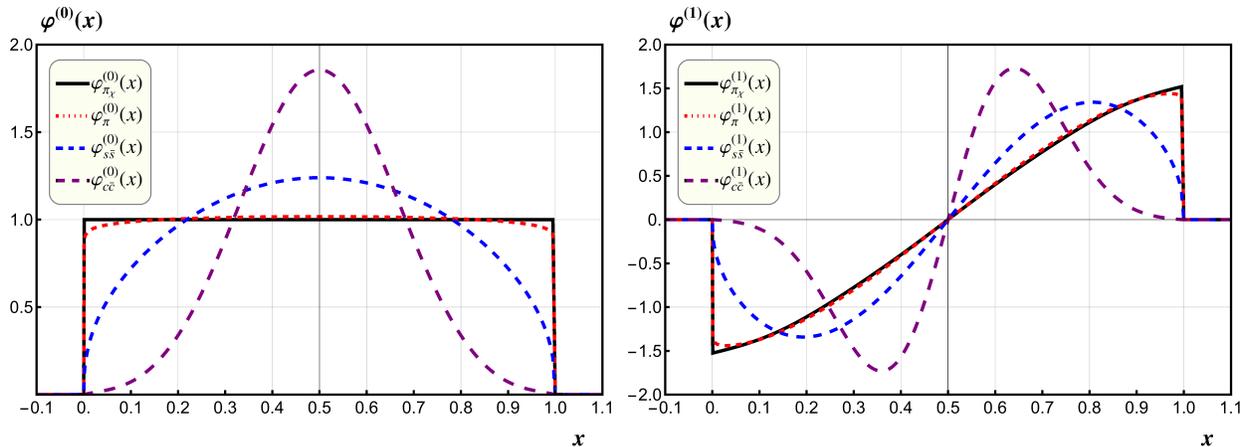}\\
\caption{The solutions to the 't Hooft (Light-front) wave functions of ground state and the first excited state mesons, for several distinct quark mass.
The ground/1st-excited state LF wave functions are even/odd under charge conjugation transformation $x\leftrightarrow 1-x$. \label{Fig:tHooft:wave:functions}}
%---------------------
\end{figure}
%---------------------

We proceed to show the profiles of various bound-state wave functions.
In Fig.~\ref{Fig:tHooft:wave:functions}, we first plot a number of
't Hooft (LF) wave functions for the ground state and the first excited state in the mass spectra,
affiliated with the different quark species.
As dictated by charge conjugation symmetry, the LF wave functions with even/odd $n$ are symmetric/antisymmetric
under the exchange $x \leftrightarrow 1-x$.
The LF wave functions always vanish in both end points $x=0,\,1$.
For lighter quark, the LF wave functions for ground states exhibit a very steep rising/falling behavior when $x$ approaches the boundaries, and a stable
plateau in the majority of range in $x$ (Note the slope in the boundaries becomes infinite in the chiral limit!).
For heavy quark, the LF wave function possesses a much milder rising/falling shape
near the end points, and the plateau disappears.

%---------------------
\begin{figure}
\centering
\includegraphics[width=\textwidth]{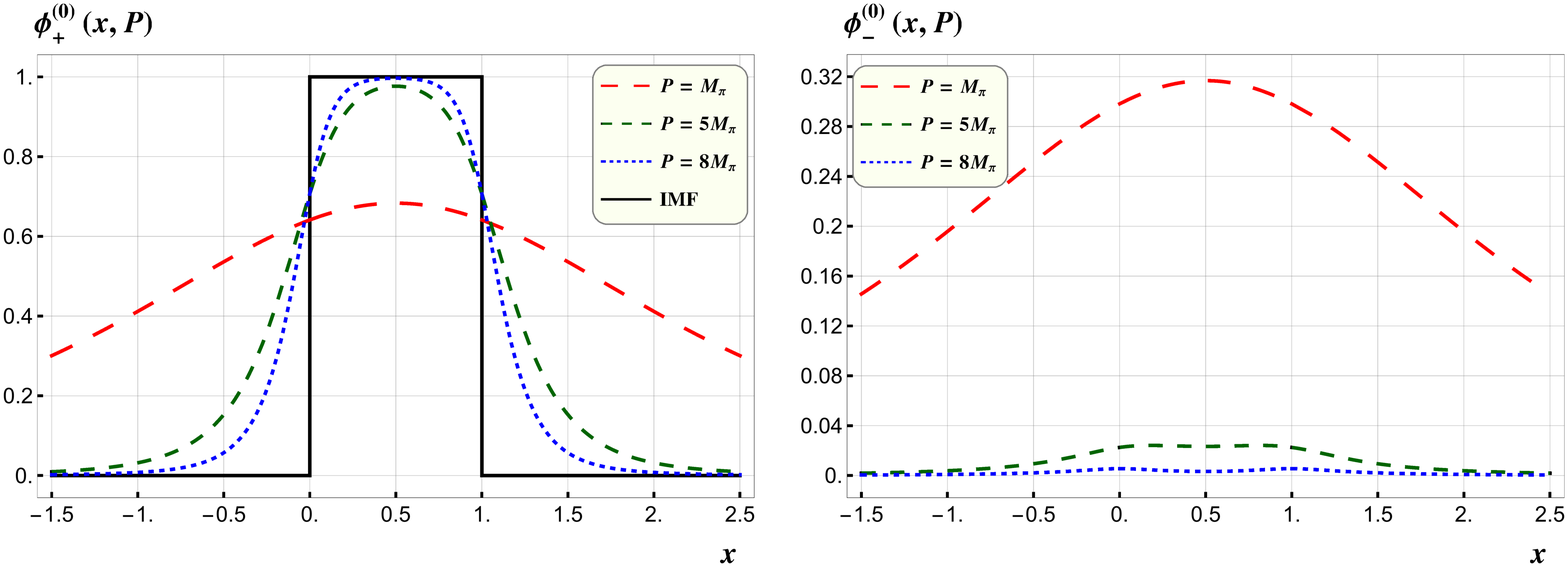}\\
\includegraphics[width=\textwidth]{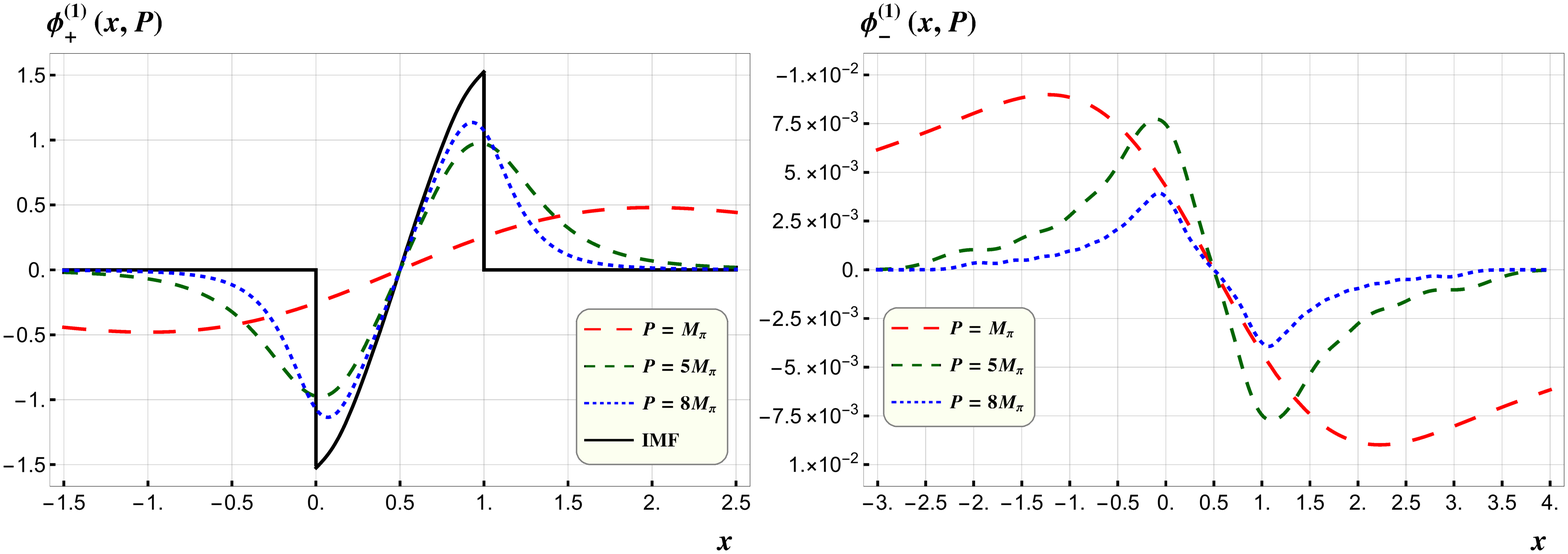}\\
\caption{Bars-Green wave functions for the low-lying $u\bar{u}$ states in the chiral limit:
chiral pion $\pi_\chi$ (first row), and the first excited state (second row). \label{Fig:BG:wave:func:massless:u}}
\end{figure}
%---------------------

%---------------------
\begin{figure}
\centering
\includegraphics[width=\textwidth]{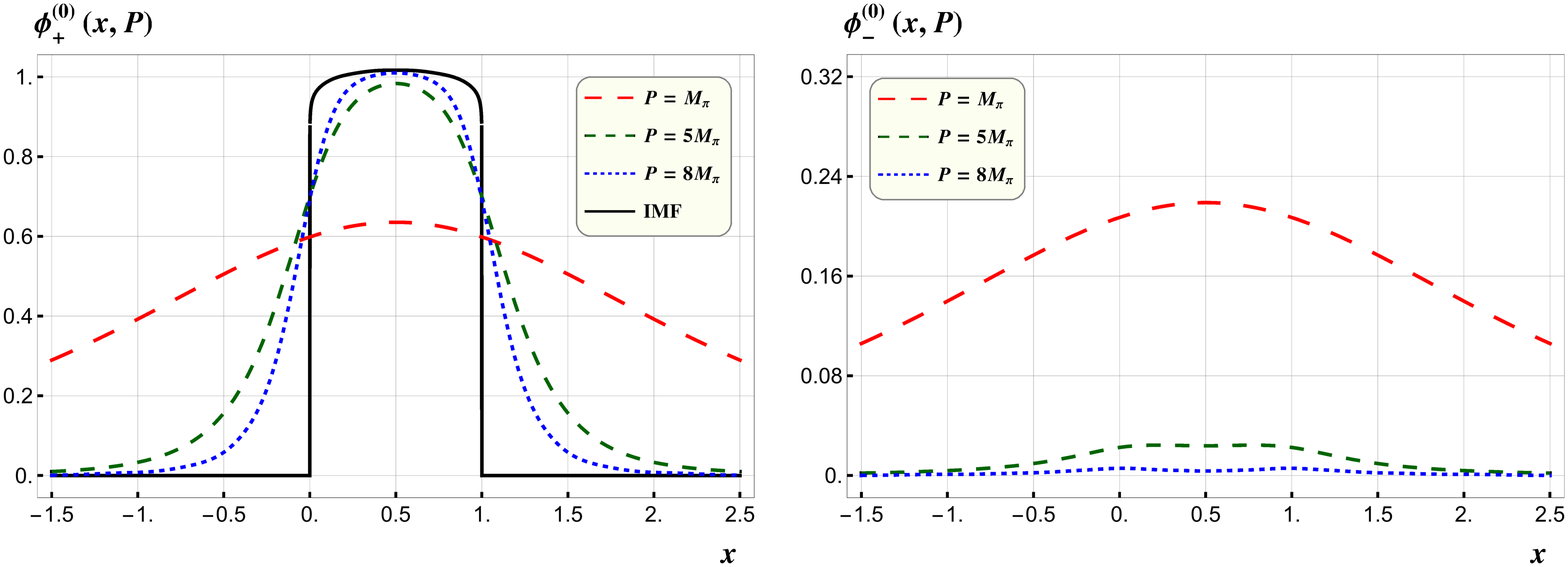}\\
\includegraphics[width=\textwidth]{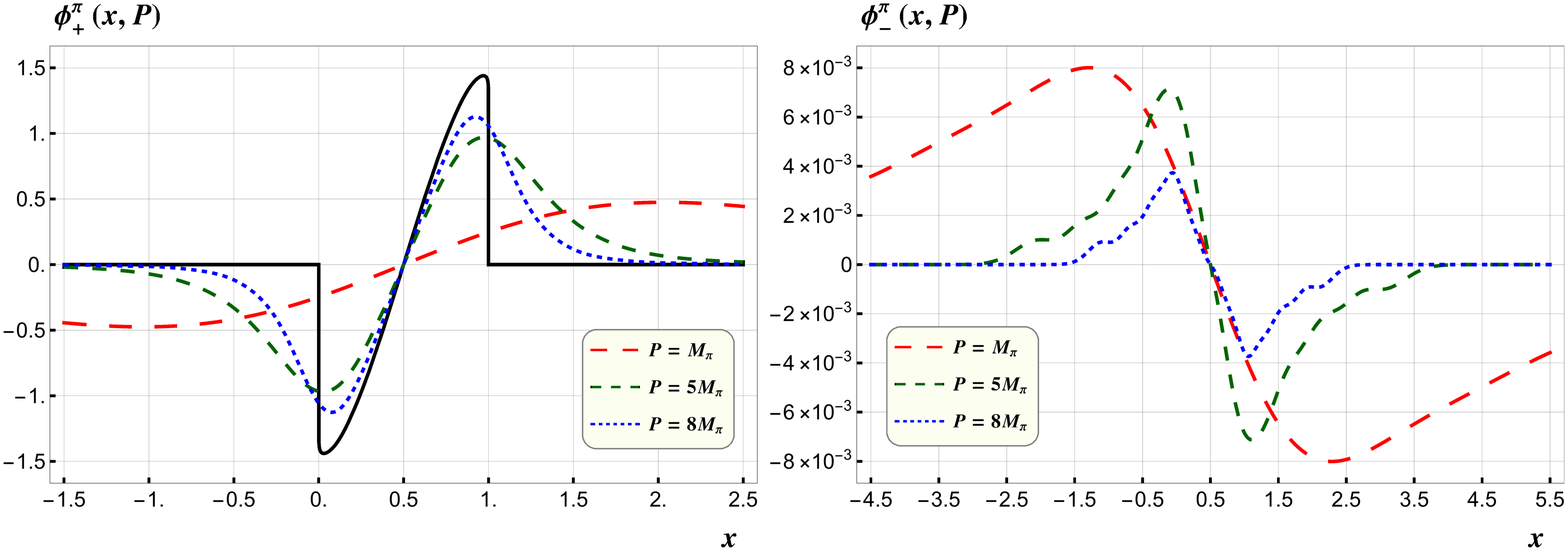}
\caption{Bars-Green wave functions for the low-lying $u\bar{u}$ states with $m_u=0.045$:
physical pion (first row), and the first excited state (second row).   \label{Fig:BG:wave:func:physical:u}}
\end{figure}
%---------------------

%---------------------
\begin{figure}
\centering
\includegraphics[width=\textwidth]{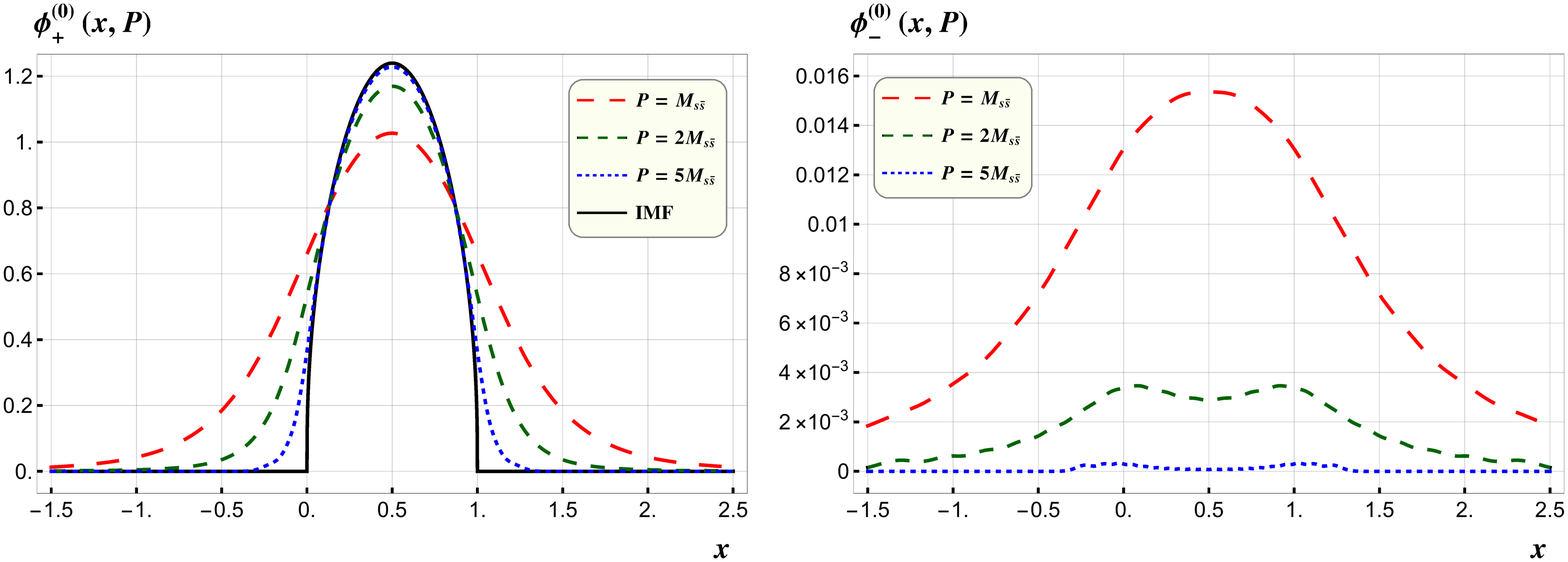}\\
\includegraphics[width=\textwidth]{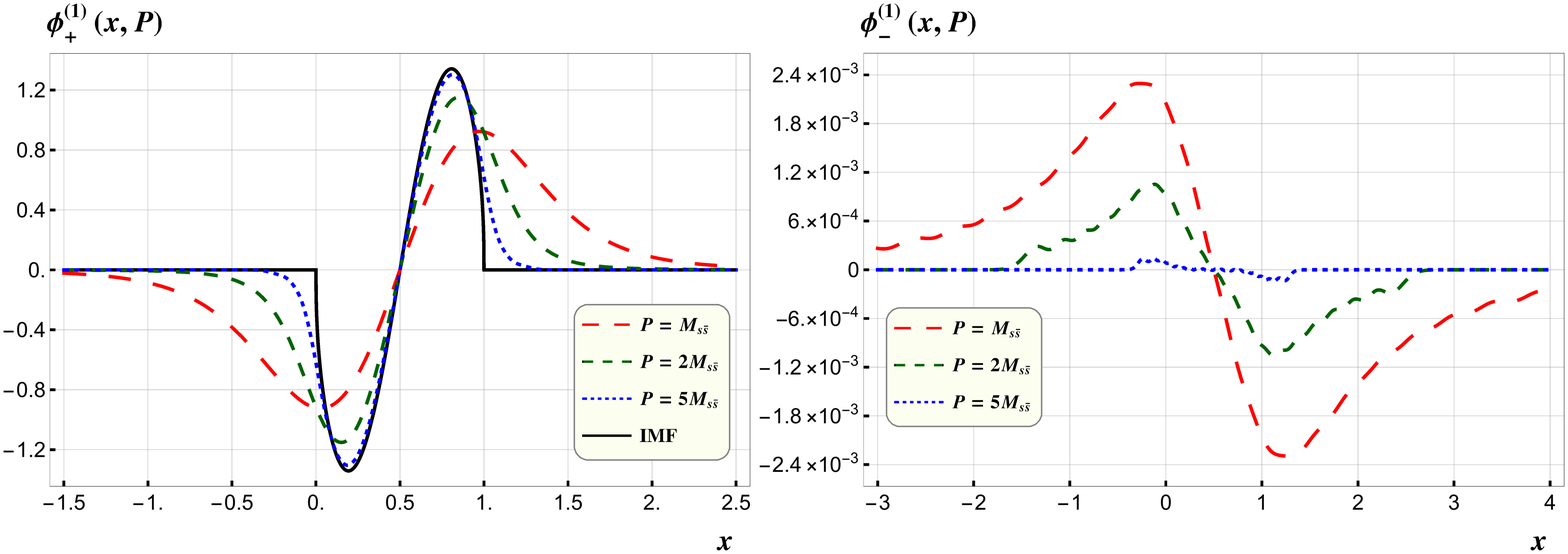}
\caption{Bars-Green wave functions for the low-lying strangeonium family with $m_s=0.749$:
Ground state (first row), and the first excited state (second row). \label{Fig:BG:wave:func:s}}
\end{figure}
%---------------------

%---------------------
\begin{figure}
\centering
\includegraphics[width=\textwidth]{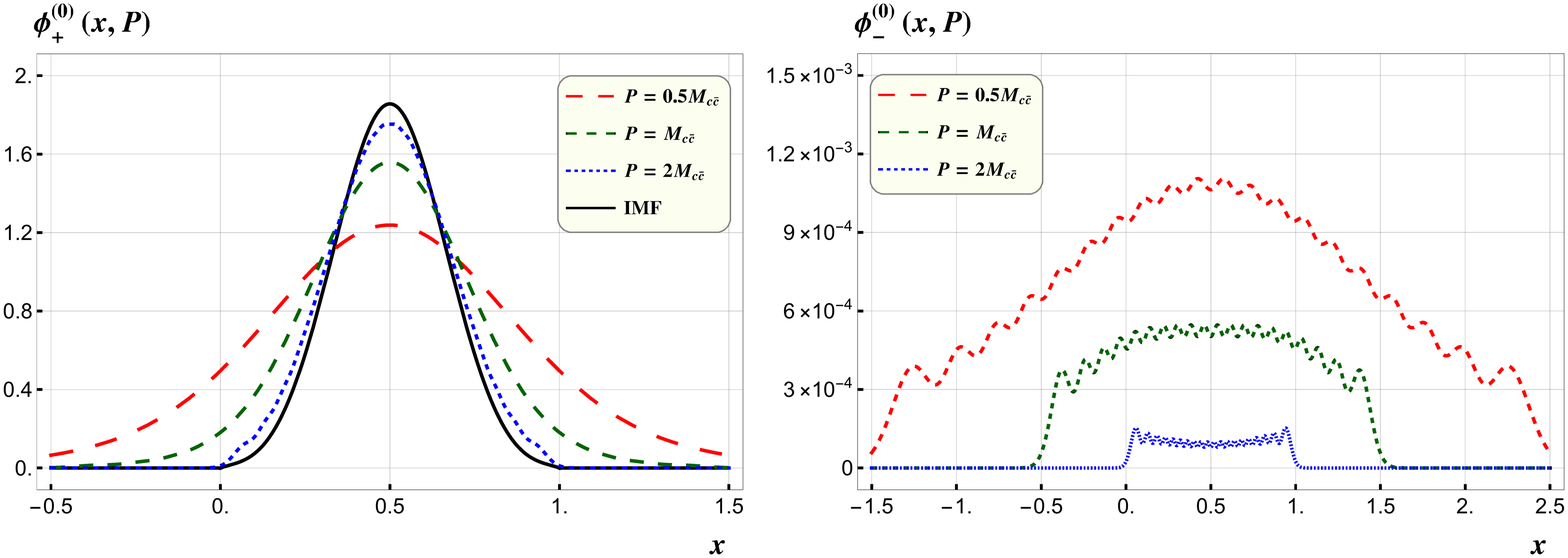}\\
\includegraphics[width=\textwidth]{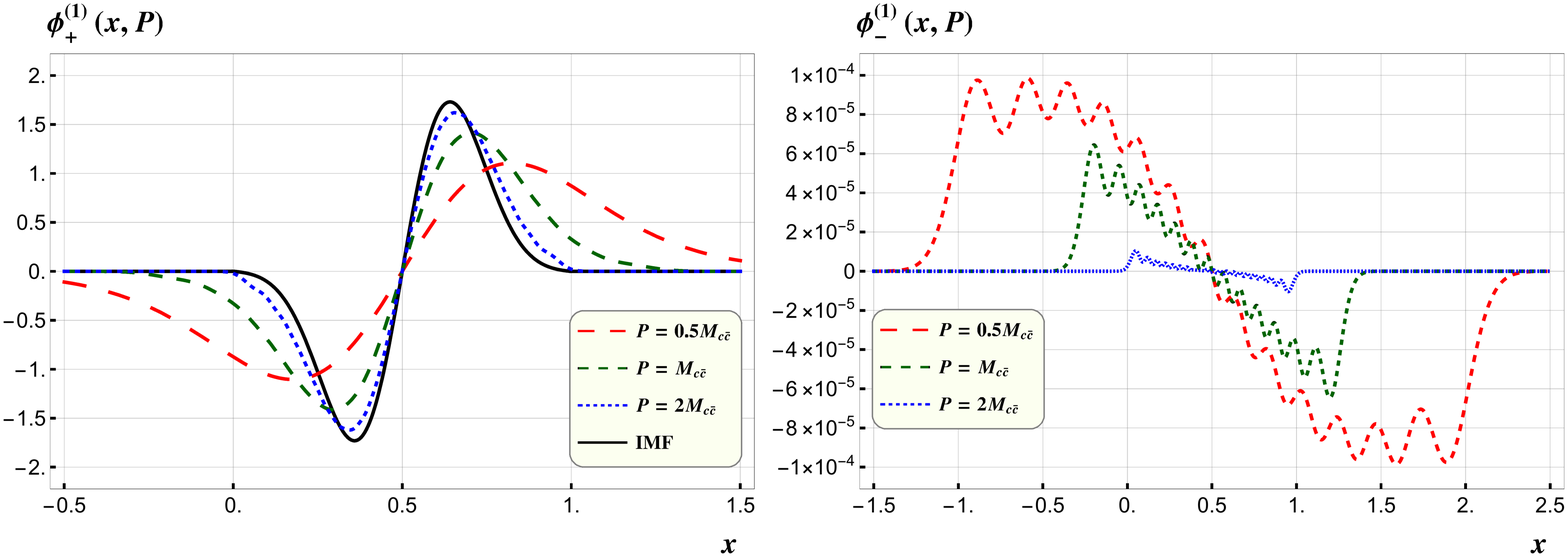}\\
\caption{Bars-Green wave functions for the low-lying charmonium states with $m_c=4.23$:
Ground state (first row), and the first excited state (second row).  \label{Fig:BG:wave:func:c}}
\end{figure}
%---------------------

In Fig.~\ref{Fig:BG:wave:func:massless:u}, \ref{Fig:BG:wave:func:physical:u}, \ref{Fig:BG:wave:func:s}
 and \ref{Fig:BG:wave:func:c}, we plot various Bars-Green wave functions of the ground state and the
first excited state
associated with several quark flavors: massless $u$, $m_u=0.045$, $m_s=0.75$, and $m_c=4.23$, respectively.
For the sake of comparison, we also juxtapose the LF wave functions of the corresponding meson in each figure.
The BG wave functions with even/odd $n$ are symmetric/antisymmetric
under the exchange $x \leftrightarrow 1-x$, as required by the charge conjugation symmetry
for flavor-neutral states.

Obviously, the BG wave functions are spatially much more spread in $x$-axis than the 't Hooft wave functions,
which are confined within the interval $x\in [0,1]$.
From these figures, one clearly observes that, for all species of quark flavors,
when the mesons are boosted with higher and higher momentum, the $\phi_+$ component of the BG wave functions
will approach the 't Hooft wave functions, while the $\phi_-$ components rapidly dies off.
These behaviors are completely compatible with the anticipated IMF limit of the BG wave functions
in (\ref{IMF:limit:Bars-Green:equation}).

In most cases, the ``backward-motion'' wave functions $\phi_-(x,P)$ are always much less significant in magnitude
than the ``forward-motion'' components
 $\phi_+(x,P)$.
The only exception is a ``wee'' (very low-momentum) lowest-lying meson,
as exemplified by chiral and physical pions in Fig.~\ref{Fig:BG:wave:func:massless:u} and Fig.~\ref{Fig:BG:wave:func:physical:u}.
As mentioned before, the comparable magnitude between $\phi_+$ and $\phi_-$ is expected for
the soft Goldstone boson, which can be intuitively attributed to
the nontrivial vacuum structure, characterized
by the nonzero quark condensate.
The $\phi_-$ component becomes quickly suppressed with respect to $\phi_+$, provided that the meson momentum
increases, or going to higher excited states, or increases the quark mass, which can be attributed to the
rapidly decreasing energy denominator
$1/(E(p)+E(P-p)+P^0)$ in (\ref{Bars:Green:equations:for:mesons}).
This is somewhat analogous to the case of the Dirac equation, where
the disparity between the large component and small component of the Dirac spinor
becomes substantial when going to nonrelativistic/ultra-relativistic limit.

We would also like to mention a technical nuisance. As can be seen in Fig.~\ref{Fig:BG:wave:func:c},
some wiggles have emerged in $\phi_-(x,P)$ for the lowest-lying and first excited charmonium state.
This should be regarded as the calculational artifact, which presumably arises from the truncation error due to
the insufficient number of our basis functions. In principle, these wiggles would vanish if we include an infinite number of orthogonal basis function. Typically in this work we choose about 20 harmonic oscillator basis functions.
Perhaps we should seek a smarter set of basis functions that allows for a faster convergence behavior.
On the other hand, we note that, whenever the wiggles appear, the corresponding $\phi_-$ component
is always typically about 3 or 4 orders-of-magnitude smaller than the $\phi_+$ component,
thus completely negligible in a practical sense.

\begin{figure}
\centering
\includegraphics[width=\linewidth]{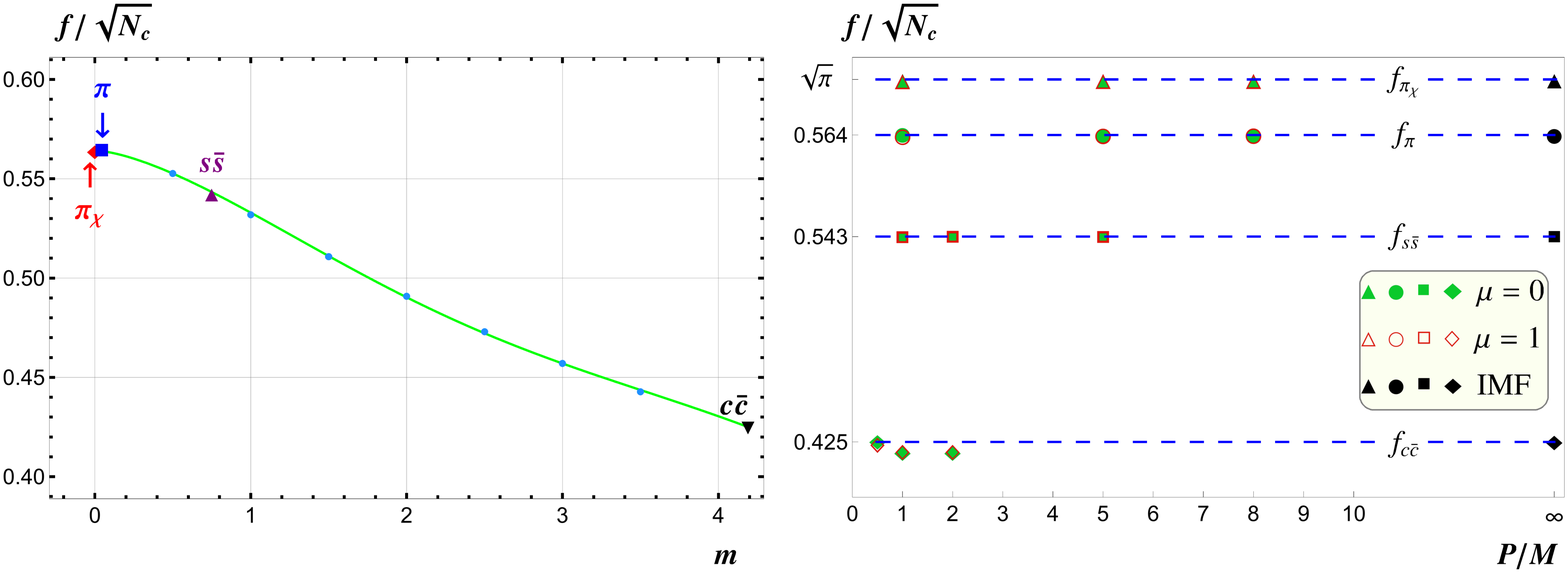}
\caption{Meson decay constants as functions of the quark mass (left) and the meson momentum (right).
 In the right panel, the colored solid/empty symbols signify
 the decay constants extracted via \ref{Decay:constant:BG:n:even} from $\mu=0, 1$ Lorentz indices, respectively,
 while the black solid symbols represent the IMF values inferred from (\ref{decay:constant:LC}).
 \label{Fig:Decay:Constants}}
\end{figure}

With the BG wave functions $\phi_\pm(x,P)$ available, we can employ the formulas derived
in Section~\ref{section:decay:constant} to compute the meson decay constant.
In Fig.~\ref{Fig:Decay:Constants}, we plot the decay constant of the ground state meson as functions of
the quark mass and meson momentum.
One finds an overall satisfactory agreement between the light-cone-gauge and axial-gauge predictions.
In the Bars-Green formalism, we have explicitly examined that the meson decay constant is
indeed frame-independent, as it must be. This can be viewed as another nontrivial verification
of the Poincar\'{e} invariance of the Bars-Green formalism.

\section{Summary}
\label{Summary}

The 't Hooft model has constantly served a fruitful theoretical laboratory
to sharpen our understanding about certain aspects of the realistic QCD.
In contrast with the widely-studied light-front quantization of ${\rm QCD}_2$,
much less work has been conducted in the equal-time quantization.
The most notable formalism in this category is based on the axial gauge quantization,
with the corresponding bound-state equations first developed by Bars and Green in late 1970s~\cite{Bars:1977ud}.
It was formally proved that when the meson is boosted to the IMF, the large component of the BG
wave equation would exactly reduce to the 't Hooft wave function.
Moreover,  it is believed that Poincar\'{e} invariance should be preserved for the color-singlet meson wave function
with arbitrary finite meson momentum. Unfortunately, until now this important feature has never
been explicitly verified in a numerical fashion.

To date, the most comprehensive numerical solutions of the BG equation were those works
done by Li and companions more than three decades ago,
yet only for the {\it stationary} mesons~\cite{Li:1986gf,Li:1987hx}.
In this paper, we have moved an important step forward,
by numerically solving the Bars-Green equation for arbitrarily {\it moving} mesons, with meson species
ranging from the chiral pion to heavy quarkonium.
We are able to numerically establish the validity of Poincar\'{e} invariance
of the 't Hooft model. Moreover, we have explicitly confirmed the tendency that, as the meson gets more and more
boosted, the large component of the Bars-Green wave function is indeed approaching the corresponding LF
wave function obtained in the light-cone gauge. We also computed the quark condensates and meson decay constants
with a variety of meson momentum, and explicitly verified the frame-independence and gauge invariance of these
physical quantities.

As a topical application, the t' Hooft model may serve as a concrete toy model
to extract some general features of the recently proposed quasi parton
distributions~\cite{Ji:2013dva}.
We note that the relation between the 't Hooft light-cone gauge formulation and the Bar-Green axial-gauge formulation for
the two-dimensional QCD, is very similar to that between the LF parton distributions and the quasi parton distributions.
Right now, the lattice simulation of the quasi distributions in the ${\rm QCD}_4$ is still in its infancy.
Therefore, we hope that our comprehensive understanding of the Bars-Green wave functions may shed some important
light on the nature of quasi-distributions in realistic QCD~\cite{Jia:2017:xx}.

\begin{acknowledgments}
%------------------------------------------------
The work of  Y.~J. and S.-R.~L. is supported in part by the National Natural Science Foundation of China under Grants
No.~11475188, No.~11261130311, No.~11621131001 (CRC110 by DGF and NSFC), by the IHEP Innovation Grant under contract number Y4545170Y2, and by the State Key Lab for Electronics and Particle Detectors.
%---------------------------------------
The work of L.-J.~L. is supported in part by the National Natural Science Foundation of China under Grants
No.~11222549 and No.~11575202.
%---------------------------------------
The work of X.-N.~X. is supported by the Deutsche Forschungsgemeinschaft (Sino-German CRC 110).
%-----------------------
\end{acknowledgments}

%-----------------------

\end{document}